\documentclass[12pt]{article}
\usepackage{amssymb,amsmath,epsfig}
\usepackage{graphicx}
\usepackage{subfig}
\usepackage[utf8]{inputenc}
\usepackage{cite}
 \allowdisplaybreaks

\begin{document}
\title{\bf Some Specific Wormhole Solutions in Extended $f(R,G,T)$ Gravity}

\author{M. Ilyas$^1$ \thanks{ilyas\_mia@yahoo.com}, A. R. Athar$^2$ \thanks{athar\_chep@hotmail.com} , Fawad Khan$^1$ \thanks{fawad.ccms@gmail.com}, Nasreen Ghafoor$^1$ \thanks{nghafoor333@gmail.com}, Haifa I. Alrebdi $^{3}$ \thanks{hialrebdi@pnu.edu.sa},\\ Kottakkaran Sooppy Nisar$^{4,5}$ \thanks{n.sooppy@psau.edu.sa}
	and Abdel-Haleem Abdel-Aty $^{6}$ \thanks{amabdelaty@ub.edu.sa}\\\\
$^1$ Institute of Physics, Gomal University,\\
Dera Ismail Khan, 29220, KP, Pakistan\\
$^2$ Institute of Physics, The Islamia University of Bahawalpur,\\
Baghdad-ul-Jadeed Campus, Bahawalpur-63100, Pakistan\\
$^3$Department of Physics, College of Science, \\Princess Nourah bint Abdulrahman University,\\ P.O. Box 84428, Riyadh 11671, Saudi Arabia\\
$^4$Department of Mathematics, College of Arts and Sciences, \\Prince Sattam bin Abdulaziz University, Wadi Aldawaser 11991, Saudi Arabia\\
$^5$Adjunct Professor, School of Technology, \\Woxsen University- Hyderabad-502345, Telangana State, India\\
$^6$Department of Physics, College of Sciences, \\University of Bisha, PO
Box 344, Bisha 61922, Saudi Arabia}

\date{}

\maketitle
\begin{abstract}
This research work provides an exhaustive investigation of the viability of different coupled wormhole (WH) geometries with the relativistic matter configurations in the $f(R,G,T)$ extended gravity framework. We consider a specific model in the context of $f(R,G,T)$-gravity for this purpose. Also, we assume a static spherically symmetric space-time geometry and a unique distribution of matter with a set of shape functions ($\beta(r)$) for analyzing different energy conditions (ECs). In addition to this, we examined WH-models in the equilibrium scenario by employing anisotropic fluid. The corresponding results are obtained using numerical methods and then presented using different plots. In this case, $f(R,G,T)$ gravity generates additional curvature quantities, which can be thought of as gravitational objects that maintain irregular WH-situations. Based on our findings, we conclude that in the absence of exotic matter, WH can exist in some specific regions of the parametric space using modified gravity model as, $f(R,G,T) = R +\alpha R^2+\beta G^n+\gamma G\ln(G)+\lambda T$.
\end{abstract}
{\bf Keywords:} $f(\mathcal{R,G,T})$-gravity; wormholes; stability; energy condition; equilibrium; anisotropic fluid.\\
{\bf PACS:} .

\section{Introduction}
The cosmic microwave background radiation (CMBR), observations of supernova of type Ia, and other discoveries have all contributed to transforming mathematical and theoretical cosmic physics \cite{ref1,ref2,ref3,ref4,ref5,ref6,ref7,ref8}. Empirical evidences from the field of cosmic physics have furnished conclusive evidence that the universe is currently experiencing an expansion of accelerated nature. Gravitational theories that are consistent with observations play a substantial role in explaining both how the universe came to be and why there are relativistic star populations. The exposure of modified gravitational theories (MGTs) developed through modifications in the Einstein-Hilbert (EH) action is the most prominent methodology to investigate cosmic acceleration. Some of the recent work in MGTs are discussed in \cite{mia1,mia2,mia3,mia4,ilyas2018charged}. Nojiri and Odintsov \cite{ref9} discussed the significance of researching MGTs in order to produce a justification for the late-time universe's expansion. Also, they gave some mathematical solutions by using the MGTs as a framework to combine a late-time accelerated universe with an inflationary universe. It is clear that MGTs use terms like $f(R)$, $f(\mathcal{T})$, and other similar terms in research papers, where $R$ and $\mathcal{T}$ stand for the Ricci scalar and the torsion scalar, respectively Refs. \cite{ref10,ref11,ref12,ref13,ref14,ref15,ref16,ref17,ref18,ref19,ref20,ref21,ref22} contain the evaluations of MGTs and dark energy (DE) that were conducted.\\ Harko \emph{et al.} \cite{ref23} proposed a generalisation of the $f(R)$-MGT using an extension in the geometric part of General Relativity (GR). They named it $f(R,T)$-MGT, while $T$ denotes trace of the ordinary energy-momentum tensor. In addition, they used the $f(R)$-MGT framework to estimate equations of motion (EoM). These EoM combine both matter and geometric parts, utilizing a fundamental metric approach. Following that, Houndjo \cite{ref24} investigated several plausible methodologies resulting from specific arrangements, such as $f(R,T) = f_{1}(R) + f_{2}(T)$, to replicate a number of cosmic solutions in the $f(R)$-MGT. He also introduced a set of scale factor possibilities that would demonstrate the universe's expansion. Various cosmic aspects are described in the Refs. \cite{ref25,ref26,ref27,ref28,ref29,ref30,ref31}, which derive from the modifications that are prescribed by two different MGTs ($f(R), f(R,T)$). These modifications include the formation and viability of star configurations; the existence of comparably large compact entities; the conformity of expansion of the universe; anisotropic and viscous solutions; the famous Raychaudhary equations; and others. M. Ilyas recently proposed a new type of MGT called $f(R,G,T)$-MGT \cite{ilyas2021compact}, while $R$ and $G$ represent the Ricci Scalar and the Gauss-Bonnet (GB) invariant, respectively. He worked in the framework of $f(R,G,T)$-MGT and explored distinct feasible models to examine various physical characteristics of cosmological objects \cite{ilyas2022energy,ilyas2023charged}.\\
The specific solutions of the Einstein gravitational field equations (FEqs) are regarded as the wormholes (WHs). The WHs act as bridges between different space-time regions. Traveling between these regions through a wormhole could take a lot less time than traveling through normal space. The black holes' (BHs) mathematical solution leads to a basis solution that is similar to the WH. Eventually, it was established that the result could be expressed as a transformation of the BH geometries with a throat (Einstein-Rosen bridge) in the middle. This bridge is a dynamic entity which is linked to the two holes and rapidly contracts to form a thin connection in between these holes. The researchers have now developed numerous WH solution; these solutions link different forms of geometry on either opening of the WH. It is an amazing characteristic of the WHs that they serve as the shortest space-time linkages; they necessarily support traveling back in time. This feature relates to the popular notion that if we could move super-luminal, we would be able to interact with the past.\\
Genetically, the WH geometries are not stable. Exclusive material(exotic matter) with negative energy density is required for stable WH geometries. This is impossible for classical matter to achieve, but quantum fluctuations in diverse domains may be capable of having it. Consequently, exotic matter is prerequisite for WH geometries in GR, such as inhomogeneous matter structures that do not meet the energy conditions (ECs), but ordinary matter does. These ECs include Null energy condition (NEC), Strong energy condition (SEC), Weak energy condition (WEC), and Dominant energy condition (DEC). It is established in literature that exotic matter (in WH throat) does not fulfill the NEC \cite{ref32,ref33,ref34,ref35,ref36,ref37}. Violating NEC is an unrealistic behavior of exotic matter, and it is appealing to reduce the use of such matter content. A significant contribution is made by Kuhfittig \cite{kuhfittig2012wormholes} to examine viable WH solutions using Einstein-Maxwell gravitation. He contended that combining geometry with usual or quintessential fluid classifications can result in a traversable-WH model. In the Einstein-Dirac-Maxwell theory domain, R A Konoplya \emph{et al.} achieved several solutions of WH in the absence of exotic matter \cite{konoplya2021traversable}. Some traversable WH solutions in the framework of GR were examined by Francisco S. N. Lobo \cite{lobo2007exotic}, using Einstein FEqs with plausible matter content. Kuhfittig \cite{kuhfittig1} examined WH model considering comparably minimum exotic matter and then further developed this view to analyze various WH geometries. FSN Lobo \cite{lobo2012wormhole} found a few WH geometries with MGT that make it possible to use a lot less exotic matter. Various WH solutions without making use of exotic matter were proposed by P Moraes \emph{et al} \cite{moraes2018nonexotic}. PK Sahoo \emph{et al.} \cite{sahoo2018phantom} discovered certain WH solutions that were compliant with the ECs and included phantom fluid. Another approach was used by Nisha Godani \emph{et al.} \cite{godani2019static,godani2020wormhole} by introducing a different $f(R,T)$ function (non-linear) to examine static traversable WH with limited strange matter. Using $f(R,T)$-MGT as a framework, Parbati Sahoo \emph{et al.} \cite{sahoo2020wormhole} suggested a novel hybrid shape function for WHs. This function indicates that there is no exotic matter present. Some models of traversable WH in the regime of traceless $f(R,T)$-MGT have been presented by Parbati Sahoo \emph{et al.} \cite{sahoo2021traversable}, eliminating the need for exotic matter. GR-theory was employed by Beato \emph{et al.} \cite{ayon2016analytic} and Canfora \emph{et al.} \cite{canfora2017topologically} to obtain an accurate, traversable, and static Lorentzian WH . This WH bears minimum couplings to a non-linear sigma model with a negative cosmic constant. They further concluded that it is not mandatory for a traversable WH to require exotic matter in the GR framework \cite{ayon2016analytic,canfora2017topologically}.\\

Multiple approaches have been adopted to investigate the theoretical foundation of WHs by studying its most plausible and realistic models. Multiple methods were considered in previous studies in order to determine the potential WH solutions, which include the inclusion of scalar fields \cite{ref38}, nonsingular space and time \cite{ref39}, and quantization effects \cite{ref40,ref41}, the use of a semi-classical gravity \cite{ref42,ref43} in brane-world gravitational background \cite{ref44,ref45}, the incorporation of Chaplygin gas along with its organized types \cite{ref46,ref47,ref48,ref49}, the use of GB-gravity \cite{ref50,ref51}, and the use of $f(T)$-MGT \cite{ref52}. Kar \cite{ref53} did research on a number of interesting properties of WHs found all over the universe. He also talked about a lot of connections between Lorentzian WH models that are not static and static WH geometries. Popov \cite{ref54} investigated the feasibility of the spherical WH geometries by incorporating $S^2\times R^2$ topology and gradually fluctuating gravitation. He took into consideration both massive and massless scalar fields in his investigation. He came to the conclusion that WH geometries are present if you can figure out the coupling parameter of curvature. Pic\'{o}n \cite{ref55} described the exotic substance by using generic interpretations of microscopic scalar field lagrangians. In the Einstein-GBg theory, Maeda and Nozawa \cite{ref56} looked at how the cosmological constant affects the way that static $n$-dimensional solutions of WH.\\

Using the MGTs framework, Lobo and Oliveira \cite{ref57} evaluated the impact of fluted exotic matter on the WH geometries. They came to the conclusion that unusual WH geometries appear as a result of the additional curvature quantities of $f(R)$, which are a component of the effective energy-momentum tensor. In addition, Garcia and Lobo \cite{ref58} introduced a range of tangible WH models by making using of the non-minimal couplings of matter-curvature. They concluded that these couplings can confine violation of the NEC at WH throat, while dealing with usual matter content. Daouda \emph{et al.} \cite{ref59} employed $f(\mathcal{T})$-MGT to construct a spherically symmetric WH solution in the presence of exotic matter content. They proposed feasible WH geometries based on the assumption that the real constant torsion scalar ($f(\mathcal{T})$) is proportional to the radial pressure component ($P_r$). In the MGTs framework, B\"{o}hmer \emph{et al.} \cite{ref60} proposed that viable WH models can obey ECs at WH throat by making particular choices for shape and $f(\mathcal{T})$ functions along with cosmic red-shift. Another research was presented by Jamil \emph{et al.} \cite{ref61} in which they examined various WH models in $f(\mathcal{T})$-MGT background. They came up with a notion that inside WH throat, the exotic matter content fulfills NEC with the additional supposition that isotropic pressure and barotropic equation of state are consistent with the geometries of WH solutions. Refs. \cite{ref61a,ref61b,ref61c} contain some feasible WH geometries in $f(R,T)$-MGT and obeying ECs. The stability of dynamical WH were studied in Ref. \cite{miady}.\\
The purpose of this study is to investigate the feasibility of static WH solutions with the $f(R,G,T)$-MGT background. This will be done without taking into account the presence of exotic matter distribution in the presence of supplementary matter. We looked at domains of the anisotropic, isotropic, and barotropic types of fluid to find out the viability of WH-models. Consequently, we examined relevant contributions from NEC, WEC, and the $\beta(r)$ in this study  The manuscript is presented as follows: $f(R,G,T)$-MGT is reviewed in section 2. The section 2 starts from the generalized form of action and includes corresponding equations of motion (EoM), incorporating WH geometrical couplings with usual matter content. In the next section, we will use the $f(R,G,T)$-MGT framework to look at some plausible WH geometries with three different fluid contents. A summary of our work and conclusive remarks are presented in the final section of this manuscript.

\section{$f(R,G,T)$ Gravitation and the Geometry of Wormhole}
The $f(R,G,T)$-MGT framework is described in this section. The framework of this theory \cite{ilyas2021compact} is an applicable generalization of the $f(R,G)$, $f(G,T)$ and $f(R,T)$-MGTs. To introduce this MGT, we first consider the EH-action in this theory as following,
\begin{equation} \label{e1}
\mathcal{S}_{f(R,G,T)}=\frac{1}{2}\kappa^{-2}\int d^{4}x\sqrt{-g}[L_{m}+f(R,G,T)],
\end{equation}
where, $L_m$ denotes the matter lagrangian. Throughout this research, we used the natural unit system, which is based on the equation $8\pi G=c=1$, where $G$ and $c$ are the constants of gravity and the speed of light, respectively. When comparing $f(R)$ and $f(R,G,T)$ MGTs, it is clear that the inclusion of different quantum effects makes $f(R,G,T)$-MGT a more viable and better choice to work in. Following that, using the variational technique with respect to metric tensor ($g_{\mu\nu}$) components, the following EoM is obtained,
\begin{align}\nonumber
&{G_{\rho \sigma }} = \\\nonumber\label{5}
&\frac{1}{{{f_{R}}\left( {{R},{G},{T}} \right)}}\left[ {{\kappa ^2}{{T}_{\rho \sigma }}} \right. - \left( {{{T}_{\rho \sigma }} + {\Theta _{\rho \sigma }}} \right){f_T}\left( {{R},{G},{T}} \right) \\\nonumber
& + \frac{1}{2}{g_{\rho \sigma }}(f\left( {{R},{G},{T}} \right) +R{f_{R}}\left( {{R},{G},{T}} \right)) + {\nabla _\rho }{\nabla _\sigma }{f_{R}}\left( {{R},{G},{T}} \right)\\\nonumber
& - {g_{\rho \sigma }}\Box{f_{R}}\left( {{R},{G},{T}} \right) - (2R{R_{\rho \sigma }} - 4R_\rho ^\xi {R_{\xi \sigma }} - 4{R_{\rho \xi \sigma \eta }}{R^{\xi \eta }}\\\nonumber
& + 2R_\rho ^{\xi \eta \delta }{R_{\sigma \xi \eta \delta }}){f_{G}}\left( {{R},{G},{T}} \right) - (2R{g_{\rho \sigma }}{\nabla ^2} - 2R{\nabla _\rho }{\nabla _\sigma } - 4{g_{\rho \sigma }}{R^{\xi \eta }}{\nabla _\xi }{\nabla _\eta }\\
& - 4{R_{\rho \sigma }}{\nabla ^2} + 4R_\rho ^\xi {\nabla _\sigma }{\nabla _\xi } + 4R_\sigma ^\xi {\nabla _\rho }{\nabla _\xi } + 4{R_{\rho \xi \sigma \eta }}{\nabla ^\xi }{\nabla ^\eta })\left. {{f_{G}}\left( {{R},{G},{T}} \right)} \right],
\end{align}
where $G_{\rho\sigma}=R_{\rho\sigma}-\frac{1}{2}g_{\rho\sigma}R$ represents the Einstein-tensor and ${\Theta _{\mu \nu }}$ is mathematically given as under,
\begin{equation}\label{2}
{\Theta _{\mu \nu }} = \frac{{{g^{\alpha \beta }}\delta {T_{\alpha
\beta }}}}{{\delta {g^{\mu \nu }}}} = {g_{\mu\nu }}{L_m} - 2{T_{\mu \nu }}- 2{g^{\alpha \beta }}\frac{{{\partial
^2}{L_m}}}{{\partial {g^{\mu \nu }}\partial {g^{\alpha \beta }}}}.
\end{equation}
We are interested in determining how the presence of anisotropic pressure influences the geometrical stainability of WH. In light of this, and keeping in mind the mathematical framework that will be presented below, we assume the relativistic source is an anisotropic matter distribution.
\begin{align}\label{eq4}
{T_{\mu \nu }} =({X_\mu}{X_\nu})\Pi+({V_\mu}{V_\nu})\left[\rho+{P_t}\right] - {P_t}{g_{\mu \nu }}.
\end{align}
In the above mathematical expression (\ref{eq4}), the symbols $\rho, P_r$ and $P_t$ denote energy density, radial pressure, and tangential pressure, respectively, while $\Pi=P_r-P_t$. Similarly, the symbols ${X_\mu}$ and ${V_\mu}$ represent the fluid's four-velocity and the radial unit four-vector, respectively  The ${X_\mu}$ and ${V_\mu}$ satisfy the requirements, ${X^\mu }{X_\mu }=-1$ and ${V^\mu }{V_\mu }=1$, if the coordinate system is co-moving. It is an appealing factor that the variation of extra force has a dependance upon the way $L_m$ holds its definitive arrangement \cite{rref36}.
If one assumes that $L_m=P$, then the extra force vanishes while $P$ represents total pressure \cite{rref37}. Parallel to the aforementioned assumption, a more comprehensive choice is $L_m=\rho$ which indicates that the extra force does not vanish \cite{rref1.1,rref1.2}. Adopting this choice, Eq.(\ref{2}) takes the following form, $${\Theta _{\mu \nu }}=- 2{T_{\mu \nu }} + \rho {g_{\mu \nu }}.$$
By using the mathematical relations we described earlier, we can write the above FEq as follows:
\begin{equation}\label{fieldequation}
{R_{\sigma \lambda }} - \frac{1}{2}R{g_{\sigma \lambda }} = T_{\sigma \lambda}^{\textit{effective}},
\end{equation}
while mathematically, $T_{\sigma \lambda}^{\textit{effective}}$ is given as,
\begin{align}\nonumber
&T_{\sigma \lambda }^{\textit{effective}} =\frac{1}{{{f_{R}}\left( {{R},{G},{T}} \right)}}\left[ {{\kappa ^2}{{T}_{\sigma \lambda }}} \right. - \left( {{{T}_{\sigma \lambda }} + {\rho g _{\sigma \lambda }}} \right){f_T}\left( {{R},{G},{T}} \right) \\\nonumber
& + \frac{1}{2}{g_{\sigma \lambda }}[f\left( {{R},{G},{T}} \right) + R{f_{R}}\left( {{R},{G},{T}} \right)] + {\nabla _\sigma }{\nabla _\lambda }{f_{R}}\left( {{R},{G},{T}} \right)\\\nonumber
& - {g_{\sigma \lambda }}\Box{f_{R}}\left( {{R},{G},{T}} \right) - (2R{R_{\sigma \lambda }} - 4R_\sigma ^\xi {R_{\xi \lambda }} - 4{R_{\sigma \xi \lambda \eta }}{R^{\xi \eta }}\\\nonumber
& + 2R_\sigma ^{\xi \eta \delta }{R_{\lambda \xi \eta \delta }}){f_{G}}\left( {{R},{G},{T}} \right) - (2R{g_{\sigma \lambda }}{\nabla ^2} - 2R{\nabla _\sigma }{\nabla _\lambda } - 4{g_{\sigma \lambda }}{R^{\xi \eta }}{\nabla _\xi }{\nabla _\eta }\\
& - 4{R_{\sigma \lambda }}{\nabla ^2} + 4R_\sigma ^\xi {\nabla _\lambda }{\nabla _\xi } + 4R_\lambda ^\xi {\nabla _\sigma }{\nabla _\xi } + 4{R_{\sigma \xi \lambda \eta }}{\nabla ^\xi }{\nabla ^\eta })\left. {{f_{G}}\left( {{R},{G},{T}} \right)} \right],
\end{align}
At this point, we are focusing on the static WHs, which by definition have spherical symmetry. The general line element in the case of static spherical geometry as follows: \cite{ref35}
\begin{equation}\label{metric1}
d{s^2} = \left[{e^{a(r)}}d{t^2} - \left\{{e^{b(r)}}d{r^2} + {r^2}\left( {d{\theta ^2} + {{\sin }^2}\theta d{\phi ^2}} \right)\right\}\right],
\end{equation}
whereas $a(r)$ signifies an arbitrary radial function, which is referred to as redshift. This arbitrary function is not constant and hence is mathematically represented as: $a(r)=-\frac{c}{r}$, where $c$ is constant. In order to mathematically describe $e^{-b}$, the $\beta(r)$ can be included in the form, ${e^{-b(r)}}=\left[{1 - \frac{{\beta(r)}}{r}}\right]$ \cite{ref35}. Corresponding to the geometry presented in Eq.(\ref{metric1}), the ${X^\mu}$ and ${V^\mu}$ are expressed as: $X^\mu=e^{-b/2}\delta^\mu_1$ and $V^\mu=e^{-a/2}\delta^\mu_0$, respectively. We are choosing a specific range for the radial coordinates, that is: $r_0$ to $\infty$, where $r_0=\beta(r_0)$. Choosing the radial coordinates in this range helps to accomplish viable configurations of the surface that agree with the WH throat. Moreover, the research establishes that satisfying the flare-out limit at the throat is a mandatory prerequisite for a WH. In this research, we choose $\frac{(\beta -\beta'r)}{\beta^2}> 0$ if the $\beta'<1$ at $r_0$. The important point here is that these conditions encourage the development of feasible WH models, even though such WH models contain exotic matter and violate NEC within the framework of GR. The matter variables ($\rho, P_r$, and $P_t$) can be fully described using the field equation (\ref{fieldequation}) in the following way, where a prime signifies a derivative w.r.t. the radial coordinate ($r$).
\begin{align}\label{rho8}\nonumber
\rho & =-\frac{1}{(2f_{R}+1)}f_{R}\left[\frac{2e^{-b}f_{RGG}\acute{G}^{2}}{f_{R}}-\frac{e^{-b}\acute{b}f_{RG}\acute{G}}{f_{R}}+\frac{4e^{-b}f_{RG}\acute{G}}{rf_{R}}+\frac{4e^{-b}\acute{T}f_{RGT}\acute{G}}{f_{R}}\right.\\\nonumber
&\left.+\frac{4e^{-b}\acute{R}f_{RRG}\acute{G}}{f_{R}}-R-GF_{G}-\frac{2e^{-b}(r\acute{b}+e^{b}-1)f_{R}}{r^{2}}-\frac{e^{-b}\acute{b}\acute{T}f_{RT}}{f_{R}}+\frac{4e^{-b}\acute{T}f_{RT}}{rf_{R}}\right.\\\nonumber
&\left.+\frac{2e^{-b}T^{\prime\prime}f_{RT}}{f_{R}}+\frac{2e^{-b}\acute{T}^{2}f_{RTT}}{f_{R}}+\frac{2e^{-b}G^{\prime\prime}f_{RG}}{f_{R}}-\frac{e^{-b}\acute{b}\acute{R}f_{RR}}{f_{R}}+\frac{4e^{-b}\acute{R}f_{RR}}{rf_{R}}\right.\\ \nonumber &\left.+\frac{2e^{-b}R^{\prime\prime}f_{RR}}{f_{R}}+\frac{4e^{-b}\acute{R}\acute{T}f_{RRT}}{f_{R}}+\frac{1}{r^24e^{-2b}}((-3+e^b)\acute{b}(\acute{T}f_{GT})+\acute{G}f_{GG})\right.\\\nonumber
&\left.+\acute{R}f_{RG}-2(-1+e^b)(f_{GGG})\acute{G}^2+2\acute{R}f_{RGG}\acute{G}+T^{\prime\prime}f_{GT}+\acute{T}^2f_{GTT}+G^{\prime\prime}f_{GG}\right.\\
&\left.+R^{\prime\prime}f_{RG}+2\acute{T}(\acute{G}f_{GGT})+\acute{R}^2f_{RRG}+\frac{2e^{-b}\acute{R}^2f_{RRR}}{f_{R}}+\frac{f}{f_{R}}\right]
\end{align}
\begin{align}\label{Pr8}\nonumber
P_{r} &
=\frac{1}{2\left(\frac{f_T+1}{f_R}+1\right)}\left(\frac{f}{f_R}-R-Gf_G-\frac{2e^{-b}(-r\acute{a}+e^b-1)f_R}{r^2}\right.\\\nonumber
&\left.-\frac{1}{r^2}4e^{-2b}(-3+e^b)\acute{a}(\acute{T}f_{GT}+\acute{G}f_{GG}+\acute{R}f_{RG})\right.\\\nonumber
&\left.+\frac{1}{rf_R}e^{-b}(r\acute{a}+4)(\acute{T}f_{RT}+\acute{G}f_{RG}+\acute{R}f_{RR})\right.\\\nonumber
&\left.+\frac{1}{(f_R+1)}f_T\left(\frac{2e^{-b}f_{RGG}\acute{G}^2}{f_R}-\frac{e^{-b}\acute{b}f_{RG}\acute{G}}{f_R}+\frac{4e^{-b}f_{RG}\acute{G}}{rf_R}+\frac{4e^{-b}\acute{T}f_{RGT}\acute{G}}{f_R}\right.\right.\\\nonumber
&\left.\left.+\frac{4e^{-b}\acute{R}f_{RRT}\acute{G}}{f_R}-R-Gf_G-\frac{2e^{-b}(r\acute{b}+e^b-1)f_R}{r^2}-\frac{e^{-b}\acute{b}\acute{T}f_{RT}}{f_R}+\frac{4e^{-b}\acute{T}f_{RT}}{rf_R}\right.\right.\\\nonumber
&\left.\left.+\frac{2e^{-b}T^{\prime\prime}f_{RT}}{f_R}+\frac{2e^{-b}\acute{T}^2f_{RTT}}{f_R}+\frac{2e^{-b}G^{\prime\prime}f_{RG}}{f_R}-\frac{e^{-b}\acute{b}\acute{R}f_{RR}}{f_R}+\frac{4e^{-b}\acute{R}f_{RR}}{rf_R}\right.\right.\\\nonumber
&\left.\left.+\frac{2e^{-b}R^{\prime\prime}f_{RR}}{f_R}+\frac{4e^{-b}\acute{R}\acute{T}f_{RRT}}{f_R}\right.\right.\\\nonumber
&\left.\left.+\frac{1}{r^2}4e^{-2b}\left((-3+e^b)\acute{b}(\acute{T}f_{GT}+\acute{G}f_{GG}+\acute{R}f_{RG})\right.\right.\right.\\\nonumber
&\left.\left.\left.-2(-1+e^b)\left(f_{GGG}\acute{G}^2+2\acute{R}f_{RGG}\acute{G}+T^{\prime\prime}f_{GT}+\acute{T}^2f_{GTT}+G^{\prime\prime}f_{GG}+R^{\prime\prime}f_{RG}\right.\right.\right.\right.\\
&\left.\left.\left.\left.+2\acute{T}(\acute{G}f_{GGT}+\acute{R}f_{RGT})+\acute{R}^2f_{RRT}\right)\right)+\frac{2e^{-b}\acute{R}^2f_{RRR}}{f_R}+\frac{f}{f_R}\right)\right)
\end{align}

\begin{align}\label{Pt8}\nonumber
P_{t} &
=\left(\frac{e^{-b}(r\acute{a}^2+(2-r\acute{b})\acute{a}-2\acute{b}+2ra^{\prime\prime})f_{R}^{2}}{r}-2Rf_R-2Gf_Gf_R\right.\\\nonumber
&\left.+\frac{1}{r}4e^{-2b}\left((\acute{T}f_{GT}+\acute{G}f_{GG}+\acute{R}f_{RG})\acute{a}^2\right.\right.\\\nonumber
&\left.\left.+\left(2\left(f_{GGG}\acute{G}^2+2\acute{R}f_{RGG}\acute{G}+T^{\prime\prime}f_{GT}+\acute{T}^2f_{GTT}+G^{\prime\prime}f_{GG}+R^{\prime\prime}f_{RG}\right.\right.\right.\right.\\\nonumber
&\left.\left.\left.\left.+2\acute{T}(\acute{G}f_{GGT}+\acute{R}f_{RGT})+\acute{R}^2f_{RRG}\right)-3\acute{b}(\acute{T}f_{GT}+\acute{G}f_{GG}+\acute{R}f_{RG})\right)\acute{a}\right.\right.\\\nonumber
&\left.\left.+2a^{\prime\prime}(\acute{T}f_{GT}+\acute{G}f_{GG}+\acute{R}f_{RG})\right)f_{R}+\frac{1}{(f_R+1)}2f_T\left(\frac{2e^{-b}f_{RGG}\acute{G}^2}{f_R}-\frac{e^{-b}\acute{b}f_{RG}\acute{G}}{f_R}\right.\right.\\\nonumber
&\left.\left.+\frac{4e^{-b}f_{RG}\acute{G}}{rf_{R}}+\frac{4e^{-b}\acute{T}f_{RGT}\acute{G}}{f_{R}}+\frac{4e^{-b}\acute{R}f_{RRG}\acute{G}}{f_{R}}-R-Gf_{G}-\frac{2e^{-b}(r\acute{b}+e^b-1)f_R}{r^2}\right.\right.\\\nonumber
&\left.\left.-\frac{e^{-b}\acute{b}\acute{T}f_{RT}}{f_R}+\frac{4e^{-b}\acute{T}f_{RT}}{rf_{R}}+\frac{2e^{-b}T^{\prime\prime}f_{RT}}{f_R}+\frac{2e^{-b}\acute{T}^2f_{RTT}}{f_R}+\frac{2e^{-b}G^{\prime\prime}f_{RG}}{f_R}\right.\right.\\\nonumber
&\left.\left.-\frac{e^{-b}\acute{b}\acute{R}f_{RR}}{f_R}+\frac{4e^{-b}\acute{R}f_{RR}}{rf_R}+\frac{2e^{-b}R^{\prime\prime}f_{RR}}{f_R}+\frac{4e^{-b}\acute{R}\acute{T}f_{RRT}}{f_R}\right.\right.\\\nonumber
&\left.\left.+\frac{1}{r^2}4e^{-2b}\left((-3+e^b)\acute{b}(\acute{T}f_{GT}+\acute{G}f_{GG}+\acute{R}f_{RG})\right.\right.\right.\\\nonumber
&\left.\left.\left.-2(-1+e^b)\left(f_{GGG}\acute{G}^2+2\acute{R}f_{RGG}\acute{G}+T^{\prime\prime}f_{GT}+\acute{T}^2f_{GTT}+G^{\prime\prime}f_{GG}+R^{\prime\prime}f_{RG}\right.\right.\right.\right.\\\nonumber
&\left.\left.\left.\left.+2\acute{T}(\acute{G}f_{GGT}+\acute{R}f_{RGT})+\acute{R}^2f_{RRG}\right)\right)+\frac{2e^{-b}\acute{R}^2f_{RRR}}{f_R}+\frac{f}{f_R}\right)f_R+2f\right.\\\nonumber
&\left.+\frac{1}{r}2e^{-b}\left(2rf_{RGG}\acute{G}^2+r\acute{a}f_{RG}\acute{G}-r\acute{b}f_{RG}\acute{G}+2f_{RG}\acute{G}+4r\acute{R}f_{RRG}\acute{G}+2rT^{\prime\prime}f_{RT}\right.\right.\\\nonumber
&\left.\left.+2r\acute{T}^2f_{RTT}+2rG^{\prime\prime}f_{RG}+r\acute{a}\acute{R}f_{RR}-r\acute{b}\acute{R}f_{RR}+2\acute{R}f_{RR}+2rR^{\prime\prime}f_{RR}\right.\right.\\
&\left.\left.+\acute{T}((r\acute{a}-r\acute{b}+2)f_{RT}+4r(\acute{G}f_{RGT}+\acute{R}f_{RRT}))+2r\acute{R}^2f_{RRR}\right)\right)/(4(f_T+f_R+1))
\end{align}
Researchers have found evidence of the medium evolution of the cores of cosmic celestial objects like galaxies and their related clusters, which shows that the interval is not linear. In order to comprehend how their structures originated, one needs to investigate their linear, quasi-linear, and comparatively linear phases. In most cases, the analytical modeling of such intricate gravity interactions is a very tedious job. As a simple alternative, numerical techniques and various viable assumptions have a significant role in obtaining solutions. Following on from the previous stages, we are now considering $f(R,G,T)$-MGT using a specific method that will be explained in the following subsections. For the sake of simplicity, we assume that $f(R,G,T)=f(R,G)+\lambda T$. As a result, the above equations can be written as,
\begin{equation}\label{rho11}
\begin{split}
\rho & =-\frac{1}{4(2\lambda+f_R+1)}\left(2\left(\frac{\lambda}{\lambda+f_R+1}+1\right)\left(2e^{-b}f_{RGG}\acute{G}^2-e^{-b}\acute{b}f_{RG}\acute{G}+\frac{4e^{-b}f_{RG}\acute{G}}{r}+4e^{-b}\acute{R}f_{RRG}\acute{G}\right.\right.\\
&\left.\left.-\frac{2e^{-b}(r\acute{b}+e^b-1)f_{R}^2}{r^2}+f-Rf_R-Gf_Gf_R+2e^{-b}G^{\prime\prime}f_{RG}-e^{-b}\acute{b}\acute{R}f_{RR}\right.\right.\\
&\left.\left.+\frac{4e^{-b}\acute{R}f_{RR}}{r}+2e^{-b}R^{\prime\prime}f_{RR}+\frac{1}{r^2}4e^{-2b}f_{R}\left((-3+e^b)\acute{b}(\acute{G}f_{GG}+\acute{R}f_{RG})-2(-1+e^b)\right.\right.\right.\\
&\left.\left.\left.(f_{GGG}\acute{G}^2+2\acute{R}f_{RGG}\acute{G}+G^{\prime\prime}f_{GG}+R^{\prime\prime}f_{RG}+\acute{R}^2f_{RRG})\right)+2e^{-b}\acute{R}^2f_{RRR}\right)\right.\\
&\left.-\frac{1}{(\lambda+f_R+1)}\lambda\left(\frac{e^{-b}(r\acute{a}^2+(2-r\acute{b})\acute{a}-2\acute{b}+2ra^{\prime\prime})f_{R}^2}{r}-2Rf_R-2Gf_Gf_R\right.\right.\\
&\left.\left.+\frac{1}{r}4e^{-2b}\left((\acute{G}f_{GG}+\acute{R}f_{RG})\acute{a}^2+\left(2\left(f_{GGG}\acute{g}^2+2\acute{R}f_{RGG}\acute{G}+G^{\prime\prime}f_{GG}+R^{\prime\prime}f_{RG}\right.\right.\right.\right.\right.\\
&\left.\left.\left.\left.\left.+\acute{R}^2f_{RRG}\right)-3\acute{b}(\acute{G}f_{GG}+\acute{R}f_{RG})\right)\acute{a}+2a^{\prime\prime}(\acute{G}f_{GG}+\acute{R}f_{RG})\right)f_R+2f\right.\right.\\
&\left.\left.+\frac{1}{r}2e^{-b}\left(2rf_{RGG}\acute{G}^2+((r\acute{a}-r\acute{b}+2)f_{RG}+4r\acute{R}f_{RRG})\acute{G}+2rG^{\prime\prime}f_{RG}+r\acute{a}\acute{R}f_{RR}\right.\right.\right.\\
&\left.\left.\left.-r\acute{b}\acute{R}f_{RR}+2\acute{R}f_{RR}+2rR^{\prime\prime}f_{RR}+2r\acute{R}^2f_{RRR}\right)\right)\right.\\
&\left.-\frac{1}{(r^2(\lambda+f_R+1))}e^{-2b}\lambda\left(\acute{a}\left(e^b\acute{R}f_{RR}r^2+2e^bf_{R}^2r-4(-3+e^b)\acute{R}f_Rf_{RG}\right.\right.\right.\\
&\left.\left.\left.+\acute{G}\left(e^br^2f_{RG}-4(-3+e^b)f_{GG}f_R\right)\right)+\acute{b}\left(e^b\acute{R}f_{RR}r^2+2e^bf_{R}^2r-4\left(-3\right.\right.\right.\right.\\
&\left.\left.\left.\left.+e^b\right)\acute{R}f_Rf_{RG}+\acute{G}(e^br^2f_{RG}-4(-3+e^b)f_{GG}f_R)\right)+2\left(-e^bR^{\prime\prime}f_{RR}r^2-e^b\acute{R}^2f_{RRR}r^2\right.\right.\right.\\
&\left.\left.\left.+4e^bR^{\prime\prime}f_Rf_{RG}-4R^{\prime\prime}f_Rf_{RG}+G^{\prime\prime}(4(-1+e^b)f_{GG}f_R-e^br^2f_{RG})+\acute{G}^2\left(4\left(-1\right.\right.\right.\right.\right.\\
&\left.\left.\left.\left.\left.+e^b\right)f_{GGG}f_R-e^br^2f_{RGG}\right)+4e^b\acute{R}^2f_Rf_{RRG}-4\acute{R}^2f_Rf_{RRG}+2\acute{G}\acute{R}\left(4(-1+e^b)f_Rf_{RGG}\right.\right.\right.\right.\\
&\left.\left.\left.\left.-e^br^2f_{RRG}\right)\right)\right)\right)
\end{split}
\end{equation}
\begin{equation}\label{Pr12}
\begin{split}
P_r &= \frac{1}{(4r^2(\lambda+f_{R}+1)(2\lambda+f_{R}+1))}\left(e^{-2b}\left(4e^bf_{R}^3-4e^{2b}f_{R}^3+4e^br\acute{a}f_{R}^3+4e^{b}f_{R}^2-4e^{2b}f_{R}^2\right.\right.\\
&\left.\left.-e^br^2\lambda\acute{a}^2f_{R}^2+8e^b\lambda f_{R}^2-8e^2b\lambda f_{R}^2+4e^br\acute{a}f_{R}^2+4e^br\lambda\acute{a}f_{R}^2+er^2\lambda\acute{a}\acute{b}f_{R}^2 \right.\right.\\
&\left.\left.-2e^br^2\lambda a^{\prime\prime}f_{R}^2-2e^{2b}r^2Gf_{G}f_{R}^2-8e^b\acute{a}\acute{G}f_{GG}f_{R}^2+24\acute{a}\acute{G}f_{GG}f_{R}^2-8e^b\acute{a}\acute{R}f_{RG}f_{R}^2\right.\right.\\
&\left.\left.+24\acute{a}\acute{R}f_{RG}f_{R}^2-2e^{2b}r^2Gf_{G}f_{R}-2e^{2b}r^2\lambda Gf_{G}f_{R}-4r\lambda\acute{a}^2\acute{G}f_{GG}f_{R}-8e^b\acute{a}\acute{G}f_{GG}f_{R}\right.\right.\\
&\left.\left.-12e^b\lambda\acute{a}\acute{G}f_{GG}f_{R}+36\lambda\acute{a}\acute{G}f_{GG}f_{R}+24\acute{a}\acute{G}f_{GG}f_{R}+4e^b\lambda\acute{b}\acute{G}f_{GG}f_{R}\right.\right.\\
&\left.\left.-12\lambda\acute{b}\acute{G}f_{GG}f_{R}+12r\lambda\acute{a}\acute{b}\acute{G}f_{GG}f_{R}-8r\lambda\acute{G}a^{\prime\prime}f_{GG}f_{R}-8e^b\lambda G^{\prime\prime}f_{GG}f_{R}+8\lambda G^{\prime\prime}f_{GG}f_{R}\right.\right.\\
&\left.\left.-8r\lambda\acute{a}G^{\prime\prime}f_{GG}f_{R}-8e^b\lambda\acute{G}^2f_{GGG}f_{R}+8\lambda\acute{G}^2f_{GGG}f_{R}-8r\lambda\acute{a}\acute{G}^2f_{GGG}f_{R}-2e^{2b}r^2R\left(\lambda\right.\right.\right.\\
&\left.\left.\left.+f_{R}+1\right)f_{R}+8e^br\acute{G}f_{RG}f_{R}+2e^br^2\acute{a}\acute{G}f_{RG}f_{R}-4r\lambda\acute{a}^2\acute{R}f_{RG}f_{R}-8e^b\acute{a}\acute{R}f_{RG}f_{R}\right.\right.\\
&\left.\left.-12e^b\lambda\acute{a}\acute{R}f_{RG}f_{R}+36\lambda\acute{a}\acute{R}f_{RG}f_{R}+24\acute{a}\acute{R}f_{RG}f_{R}+4e^b\lambda\acute{b}\acute{R}f_{RG}f_{R}\right.\right.\\
&\left.\left.-12\lambda\acute{b}\acute{R}f_{RG}f_{R}+12r\lambda\acute{a}\acute{b}\acute{R}f_{RG}f_{R}-8r\lambda\acute{R}a^{\prime\prime}f_{RG}f_{R}-8e^b\lambda R^{\prime\prime}f_{RG}f_{R}+8\lambda R^{\prime\prime}f_{RG}f_{R}\right.\right.\\
&\left.\left.-8r\lambda\acute{a}R^{\prime\prime}f_{RG}f_{R}-16e^b\lambda\acute{G}\acute{R}f_{RGG}f_{R}+16\lambda\acute{G}\acute{R}f_{RGG}f_{R}-16r\lambda\acute{a}\acute{G}f_{RGG}f_{R}\right.\right.\\
&\left.\left.+8e^br\acute{R}f_{RR}f_{R}+2e^br^2\acute{a}\acute{R}f_{RR}f_{R}-8e^b\lambda\acute{R}^2f_{RRG}f_{R}+8\lambda\acute{R}^2f_{RRG}f_{R}\right.\right.\\
&\left.\left.-8r\lambda\acute{a}\acute{R}^{2}f_{RRG}f_{R}+2e^{2b}r^2f(\lambda+f_{R}+1)+8e^{b}r\acute{G}f_{RG}+12e^br\lambda\acute{G}f_{RG}\right.\right.\\
&\left.\left.+2e^br^2\acute{a}\acute{G}f_{RG}+e^br^2\lambda\acute{a}\acute{G}f_{RG}+e^br^2\lambda\acute{b}\acute{G}f_{RG}-2e^br^2\lambda G^{\prime\prime}f_{RG}-2e^br^2\lambda\acute{G}^2f_{RGG}\right.\right.\\
&\left.\left.+8e^br\acute{R}f_{RR}+12e^br\lambda\acute{R}f_{RR}+2e^br^2\acute{a}\acute{R}f_{RR}+e^br^2\lambda\acute{a}\acute{R}f_{RR}+e^br^2\lambda\acute{b}\acute{R}f_{RR}\right.\right.\\
&\left.\left.-2e^br^2\lambda R^{\prime\prime}f_{RR}-4e^br^2\lambda\acute{G}\acute{R}f_{RRG}-2e^br^2\lambda\acute{R}^2f_{RRR}\right)\right)
\end{split}
\end{equation}
\begin{equation}\label{Pt13}
\begin{split}
P_{t}&=\frac{1}{(4r^2(\lambda+f_{R}+1)(2\lambda+f_{R}+1))}\left(e^{-2b}\left(e^br^2\acute{a}^2f_{R}^3+2e^{b}r\acute{a}f_{R}^3-2e^{b}r\acute{b}f_{R}^3-e^{b}r^2\acute{a}\acute{b}f_{R}^3\right.\right.\\
&\left.\left.+2e^{b}r^{2}a^{\prime\prime}f_{R}^3+e^{b}r^{2}\acute{a}^{2}f_{R}^2+e^{b}r^{2}\lambda\acute{a}^{2}f_{R}^2+2e^{b}r\acute{a}f_{R}^2-2e^{b}r\acute{b}f_{R}^2-4e^{b}r\lambda \acute{b}f_{R}^2\right.\right.\\
&\left.\left.-e^{b}r^{2}\acute{a}\acute{b}f_{R}^2-e^{b}r^{2}\lambda\acute{a}\acute{b}f_{R}^2+2e^{b}r^{2}a^{\prime\prime}f_{R}^2+2e^{b}r^{2}\lambda a^{\prime\prime}f_{R}^2-2e^{2b}r^{2}Gf_{G}f_{R}^2\right.\right.\\
&\left.\left.+4r\acute{a}^2\acute{G}f_{GG}f_{R}^2-12r\acute{a}\acute{b}\acute{G}f_{GG}f_{R}^2+8r\acute{G}a{\prime\prime}f_{GG}f_{R}^2+8r\acute{a}G^{\prime\prime}f_{GG}f_{R}^2\right.\right.\\
&\left.\left.+8r\acute{a}\acute{G}^2f_{GGG}f_{R}^2+4r\acute{a}^2\acute{R}f_{RG}f_{R}^2-12r\acute{a}\acute{b}\acute{R}f_{RG}f_{R}^2+8r\acute{R}a^{\prime\prime}f_{RG}f_{R}^2\right.\right.\\
&\left.\left.+8r\acute{a}R^{\prime\prime}f_{RG}f_{R}^2+16r\acute{a}\acute{G}\acute{R}f_{RGG}f_{R}^2+8r\acute{a}\acute{R}^2f_{RRG}f_{R}^2-2e^{2b}r^2Gf_{G}f_{R}\right.\right.\\
&\left.\left.-2e^{2b}r^2\lambda Gf_{G}f_{R}+4r\acute{a}^2\acute{G}f_{GG}f_{R}+4r\lambda\acute{a}^2\acute{G}f_{GG}f_{R}+4e^b\lambda\acute{a}\acute{G}f_{GG}f_{R}\right.\right.\\
&\left.\left.-12\lambda\acute{a}\acute{G}f_{GG}f_{R}+4e^b\lambda\acute{b}\acute{G}f_{GG}f_{R}-12\lambda\acute{b}\acute{G}f_{GG}f_{R}-12r\acute{a}\acute{b}\acute{G}f_{GG}f_{R}\right.\right.\\
&\left.\left.-12r\lambda\acute{a}\acute{b}\acute{G}f_{GG}f_{R}+8r\acute{G}a^{\prime\prime}f_{GG}f_{R}+8r\lambda\acute{G}a^{\prime\prime}f_{GG}f_{R}-8e^b\lambda G^{\prime\prime}f_{GG}f_{R}+8\lambda G^{\prime\prime}f_{GG}f_{R}\right.\right.\\
&\left.\left.+8r\acute{a}G^{\prime\prime}f_{GG}f_{R}+8r\lambda\acute{a}G^{\prime\prime}f_{GG}f_{R}-8e^b\lambda\acute{G}^2f_{GGG}f_{R}+8\lambda\acute{G}^2f_{GGG}f_{R}+8r\acute{a}\acute{G}^2f_{GGG}f_{R}\right.\right.\\
&\left.\left.+8r\lambda\acute{a}\acute{G}^2f_{GGG}f_{R}-2e^{2b}r^2R(\lambda+f_{R}+1)f_{R}+4e^br\acute{G}f_{RG}f_{R}+2e^br^2\acute{a}\acute{G}f_{RG}f_{R}\right.\right.\\
&\left.\left.-2e^br^2\acute{b}\acute{G}f_{RG}f_{R}+4r\acute{a}^2\acute{R}f_{RG}f_{R}+4r\lambda\acute{a}^2\acute{R}f_{RG}f_{R}+4e^b\lambda\acute{a}\acute{R}f_{RG}f_{R}\right.\right.\\
&\left.\left.-12\lambda\acute{a}\acute{R}f_{RG}f_{R}+4e^b\lambda\acute{b}\acute{R}f_{RG}f_{R}-12\lambda\acute{b}\acute{R}f_{RG}f_{R}-12r\acute{a}\acute{b}\acute{R}f_{RG}f_{R}\right.\right.\\
&\left.\left.-12r\lambda\acute{a}\acute{b}\acute{R}f_{RG}f_{R}+8r\acute{R}a^{\prime\prime}f_{RG}f_{R}+8r\lambda\acute{R}a^{\prime\prime}f_{RG}f_{R}+4e^br^2G^{\prime\prime}f_{RG}f_{R}\right.\right.\\
&\left.\left.-8e^b\lambda R^{\prime\prime}f_{RG}f_{R}+8\lambda R^{\prime\prime}f_{RG}f_{R}+8r\acute{a}R^{\prime\prime}f_{RG}f_{R}+8r\lambda\acute{a}R^{\prime\prime}f_{RG}f_{R}+4e^br^2\acute{G}^2f_{RGG}f_{R}\right.\right.\\
&\left.\left.-16e^b\lambda\acute{G}\acute{R}f_{RGG}f_{R}+16\lambda\acute{G}\acute{R}f_{RGG}f_{R}+16r\acute{a}\acute{G}\acute{R}f_{RGG}f_{R}+16r\lambda\acute{a}\acute{G}\acute{R}f_{RGG}f_{R}\right.\right.\\
&\left.\left.+4e^br\acute{R}f_{RR}f_{R}+2e^br^2\acute{a}\acute{R}f_{RR}f_{R}-2e^br^2\acute{b}\acute{R}f_{RR}f_{R}+4e^br^2R^{\prime\prime}f_{RR}f_{R}\right.\right.\\
&\left.\left.-8e^b\lambda\acute{R}^2f_{RRG}f_{R}+8\lambda\acute{R}^2f_{RRG}f_{R}+8r\acute{a}\acute{R}^2f_{RRG}f_{R}+8r\lambda\acute{a}\acute{R}^2f_{RRG}f_{R}\right.\right.\\
&\left.\left.+8e^br^2\acute{G}\acute{R}f_{RRG}f_{R}+4e^br^2\acute{R}^2f_{RRR}f_{R}+2e^{2b}r^2f(\lambda+f_R+1)+4e^br\acute{G}f_{RG}\right.\right.\\
&\left.\left.+4e^br\lambda\acute{G}f_{RG}+2e^br^2\acute{a}\acute{G}f_{RG}+e^br^2\lambda\acute{a}\acute{G}f_{RG}-2e^br^2\acute{b}\acute{G}f_{RG}-3e^br^2\lambda\acute{b}\acute{G}f_{RG}\right.\right.\\
&\left.\left.+4e^br^2G^{\prime\prime}f_{RG}+6e^br^2\lambda G^{\prime\prime}f_{RG}+4e^br^2\acute{G}^2f_{RGG}+6e^br^2\lambda\acute{G}^2f_{RGG}+4e^br\acute{R}f_{RR}\right.\right.\\
&\left.\left.+4e^br\lambda\acute{R}f_{RR}+2e^br^2\acute{a}\acute{R}f_{RR}+e^br^2\lambda\acute{a}\acute{R}f_{RR}-2e^br^2\acute{b}\acute{R}f_{RR}-3e^br^2\lambda\acute{b}\acute{R}f_{RR}\right.\right.\\
&\left.\left.+4e^br^2R^{\prime\prime}f_{RR}+6e^br^2\lambda R^{\prime\prime}f_{RR}+8e^br^2\acute{G}\acute{R}f_{RRG}+12e^br^2\lambda\acute{G}\acute{R}f_{RRG}\right.\right.\\
&\left.\left.+4e^br^2\acute{R}^2f_{RRR}+6e^br^2\lambda\acute{R}^2f_{RRR}\right)\right)
\end{split}
\end{equation}

The Raychaudhuri equations are the basis for discussing ECs. The ECs play a vital role in studying the physical and realistic configuration of matter. The aforesaid conditions are applied to the energy-momentum tensor and have the festinating aspect of coordinate invariance. The Raychaudhuri equations explains how the expansion scalar ($\theta$) evolves (for congruences of null $l_{\alpha}$ and time-like $v^{\alpha}$ geodesics) w.r.t temporal coordinates as \cite{poisson2004relativist},
\begin{equation*}
 R_{\alpha\beta}v^{\alpha}v^{\beta}+\frac{\theta^2}{3}+\sigma^{\alpha\beta}\sigma_{\alpha\beta}-\omega^{\alpha\beta}\omega_{\alpha\beta}+\frac{d\theta}{d\tau}=0,
\end{equation*}
\begin{equation*}
 R_{\alpha\beta}l^{\alpha}l^{\beta}+\frac{\theta^2}{2}+\sigma^{\alpha\beta}\sigma_{\alpha\beta}-\omega^{\alpha\beta}\omega_{\alpha\beta}+\frac{d\theta}{d\tau}=0,
\end{equation*}\\
while the symbol $\sigma^{\alpha\beta}$ denotes shear tensor and the symbol $\omega^{\alpha\beta}$ denotes rotation tensor. The $\theta$ evolves w.r.t. temporal coordinates, taking into account the acceleration term while dealing with non-geodesic null congruence or time-like congruence. The following expression shows this evolution of $\theta$ \cite{dadhich2005derivation}
\begin{equation}\label{shareef13}
  R_{\alpha\beta}v^{\alpha}v^{\beta}+\frac{\theta^2}{3}+\sigma^{\alpha\beta}\sigma_{\alpha\beta}-\omega^{\alpha\beta}\omega_{\alpha\beta}-\mathcal{B}+\frac{d\theta}{d\tau}=0,
\end{equation}
here $\mathcal{B}$ is the auxiliary term which is given as, $\mathcal{B}=\nabla_\alpha\left[u^\beta\nabla_\beta u^\alpha\right]$ (four-acceleration divergence). In Eq. \ref{shareef13}, the $\mathcal{B}$ denotes the acceleration term. The pressure gradient (non-gravity force) causes this acceleration term. In the case of non-geodesic congruences, if we exclude quadratic terms and consider that gravity is attractive ($\theta<0$), the Raychaudhuri equations can be written as follows:
\begin{equation*}
  R_{\alpha\beta}l^{\alpha}l^{\beta}-\mathcal{B}\geq0, \qquad R_{\alpha\beta}v^{\alpha}v^{\beta}-\mathcal{B}\geq0.
\end{equation*}
These inequalities can also be written in relation to energy-momentum tensor as,
\begin{equation}\label{shareef14}
 \left[T_{\alpha\beta}-\frac{g_{\alpha\beta}T}{2}\right]l^{\alpha}l^{\beta}-\mathcal{B}\geq0, \qquad \left[T_{\alpha\beta}-\frac{g_{\alpha\beta}T}{2}\right]v^{\alpha}v^{\beta}-\mathcal{B}\geq0.
\end{equation}
As a result of the essentially geometric structure of the Raychaudhuri equations, the inequalities (\ref{shareef14}) can be established in MGTs using $T_{\alpha\beta}^{eff}$ instead of $T_{\alpha\beta}$. Using $T_{\alpha\beta}^{eff}$ in aforementioned inequalities (\ref{shareef14}) furnishes NEC, WEC, SEC, and DEC as,
\begin{itemize}
  \item NEC: $\rho^{eff}+P_{j}^{eff}-\mathcal{B}\geq0, \qquad j=r,t$
  \item WEC: $\rho^{eff}+P_{j}^{eff}-\mathcal{B}\geq0, \qquad \rho^{eff}-\mathcal{B}\geq0,$
  \item SEC: $\rho^{eff}+P_{j}^{eff}-\mathcal{B}\geq0, \qquad \rho^{eff}+\sum_{j}P_{j}^{eff}-\mathcal{B}\geq0,$
  \item DEC: $\rho^{eff}\pm P_{j}^{eff}-\mathcal{B}\geq0, \qquad \rho^{eff}-\mathcal{B}\geq0$,
\end{itemize}
It is generally agreed that the NEC is the most fundamental constraint. Unfulfillment of the NEC results in violation of other ECs as well. In case of non-local gravity, the Ecs were studied in Ref. \cite{mianon}. The important point here is that in the GR framework, one can get non-geodesic ECs by incorporating $\rho$ and $P$ instead of $\rho^{eff}$ and $P^{eff}$.\\
The fundamental constraint on the viability of traversable WH is the violation of the NEC. By violating NEC, throat of the WH does not contract, so the unrealistic physical solution of the WH is reached. In the regime of MGTs, the $T_{\alpha\beta}^{eff}$ serves as a substitute to fulfill this violation constraint. This characteristic of MGTs provides an advantage for the ordinary distribution of matter to obey the ECs. In our case we obtain the following NEC in $f(R,G,T)$-MGT,
\begin{equation}
\rho^{eff}+P_r^{eff}=r^{-3}\left[\left(r \beta '-\beta \right) f_R (R,G)\right]
\end{equation}

\subsection{Specific Model and Anisotropic Matter Distribution}
We are investigating the following specific model within the context of the $f(R,G,T)$-MGT:
\begin{equation}\label{cmodel}
f(R,G,T) = R +\alpha R^2+\beta G^n+\gamma G\ln(G)+\lambda T.
\end{equation}
In the above considered model (\ref{cmodel}), the Greek letters $\alpha,\beta, \gamma, \lambda$ and $n$ represent different constant parameters of the model. We are incorporating $G^n$ (GB-curvature scalar) in our model with $n$ being a real constant that includes a scale invariant FEq. It is assumed to be one of the possible and equivalent approaches to $f(R)$-MGTs. A constant, $\beta$ is multiplying with $G^{n}$ for parametrization. Moreover, there is a limitation that the logarithmic term in this model should be dimensionless. We may choose this term as $\log\left[\frac{G}{G_0}\right]$. Because the $\beta$-terms in the FEq are insignificant, the change in $G_0$ value can be adjusted or avoided by reconstructing $\beta$. Using classical approaches, it is likely to set $\beta$-terms equivalent to zero.

\begin{figure} \centering
\epsfig{file=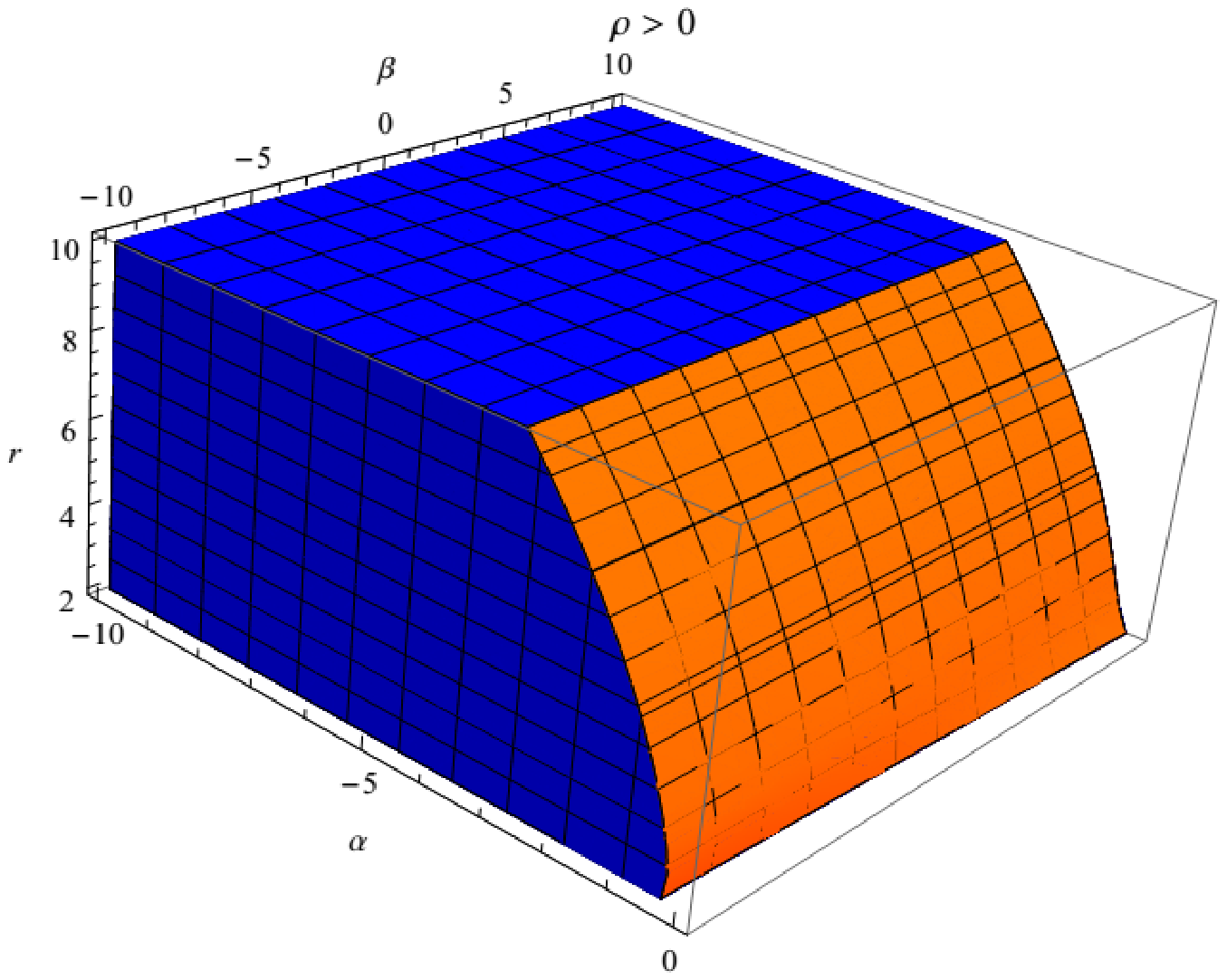,width=.48\linewidth}
\epsfig{file=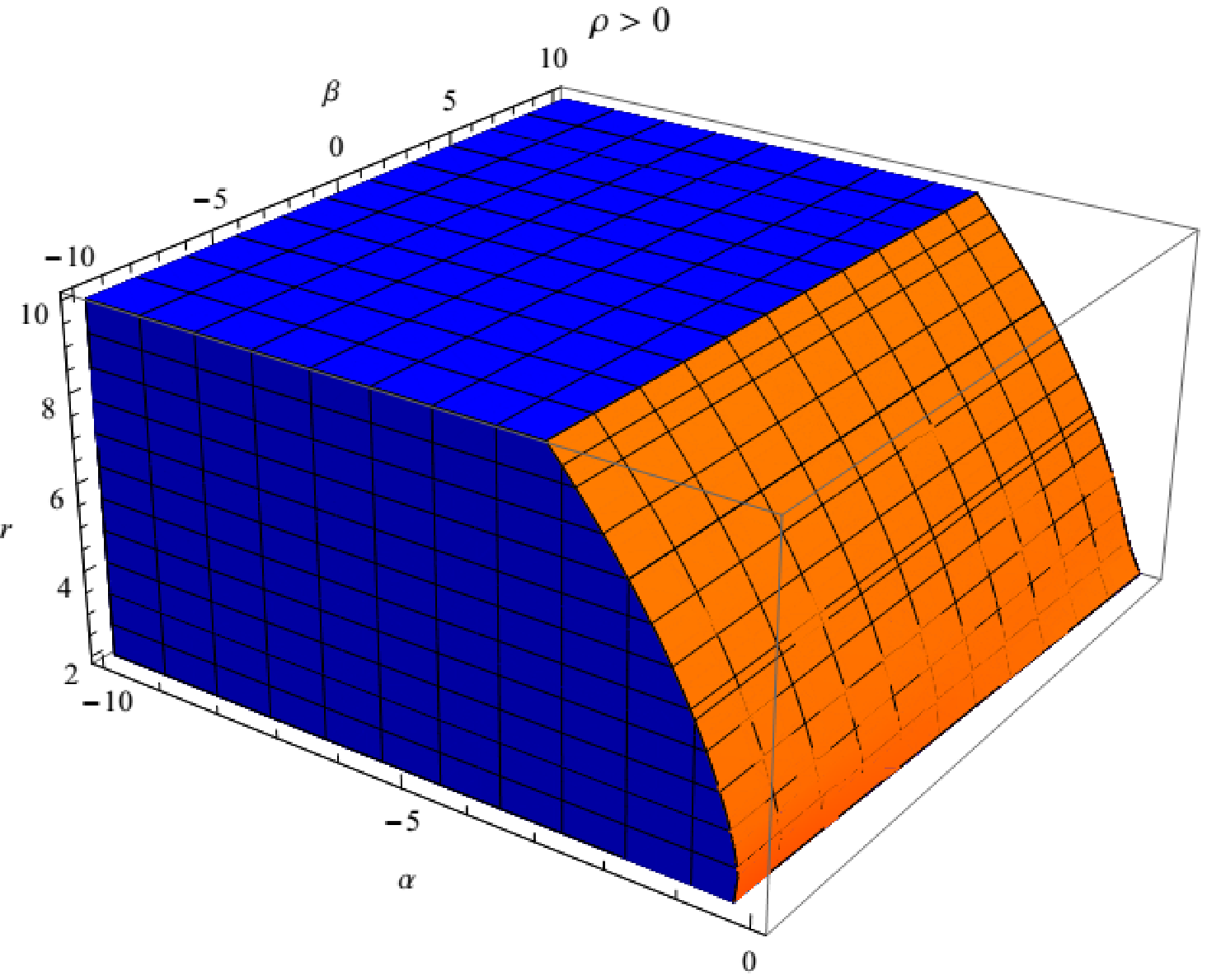,width=.48\linewidth}
\caption{The evolution of $\rho$ w.r.t. $r$, with the left side representing $\lambda>0$ and the right side representing $\lambda<0$.} \label{f1}
\end{figure}

\begin{itemize}
  \item For the validity of $\rho \geq 0$, as shown in Fig. (\ref{f1}).\\

  we see that $r\rightarrow 9$, the other parameter should be $\alpha\rightarrow -9.701$ and $\beta\rightarrow -9.50$ while for $r\rightarrow 0.1$ then $\alpha\rightarrow 5.30$ and $\beta\rightarrow 9.99$.\\

For $\alpha\rightarrow 10$, the other parameter should be $r\rightarrow 9$ and $\beta\rightarrow 10$ while for $\alpha\rightarrow -10$ then $r\rightarrow 8.87$ and $\beta\rightarrow 1.49$.\\

For $\beta\rightarrow 10$, the other parameter should be $\alpha\rightarrow -7.81$ and $r\rightarrow 1.83$ while for $\beta\rightarrow -10$ then $\alpha\rightarrow -5.62$ and $r\rightarrow 1.92$.\\
\end{itemize}

\begin{figure} \centering
\epsfig{file=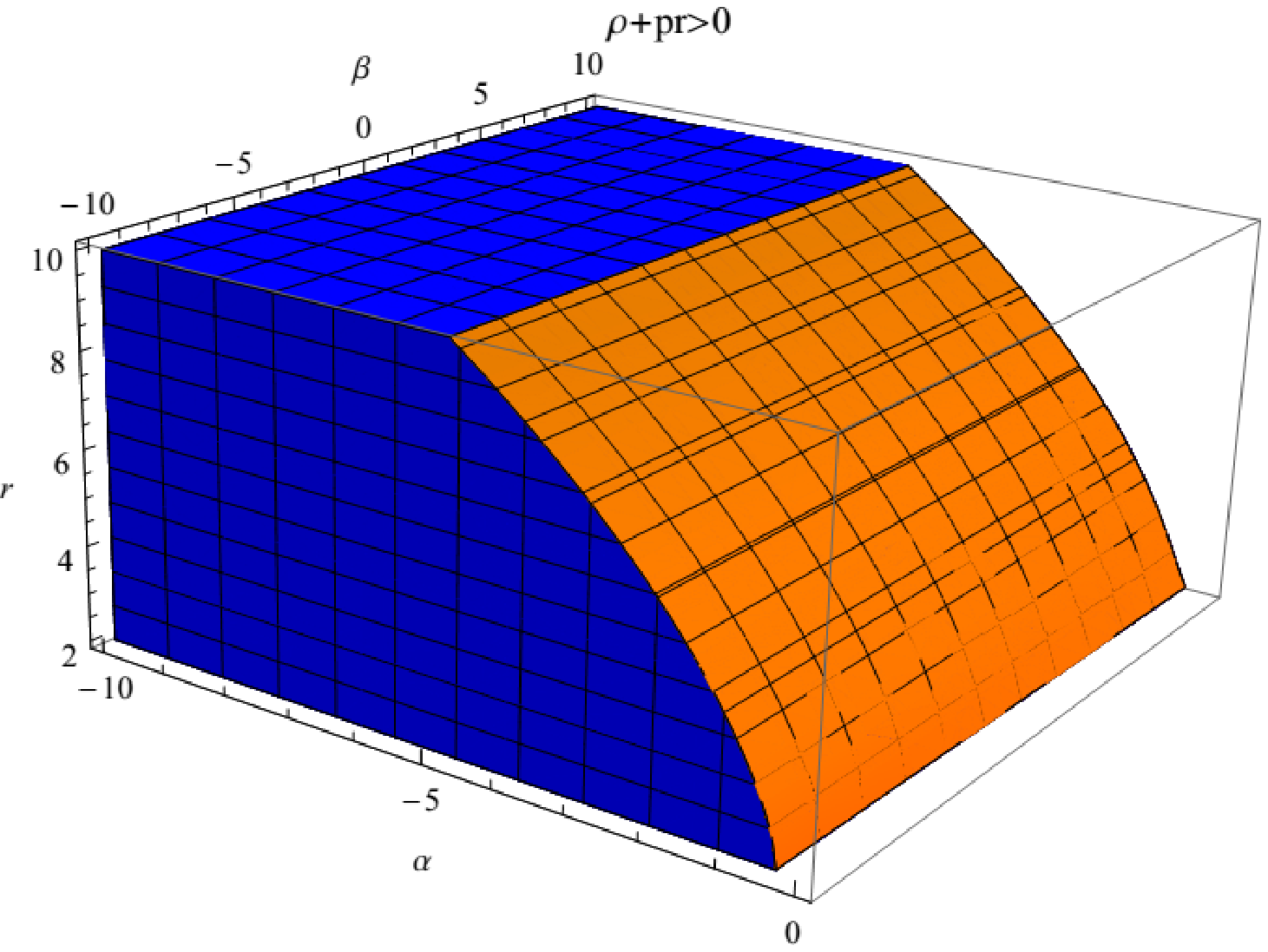,width=.48\linewidth}
\epsfig{file=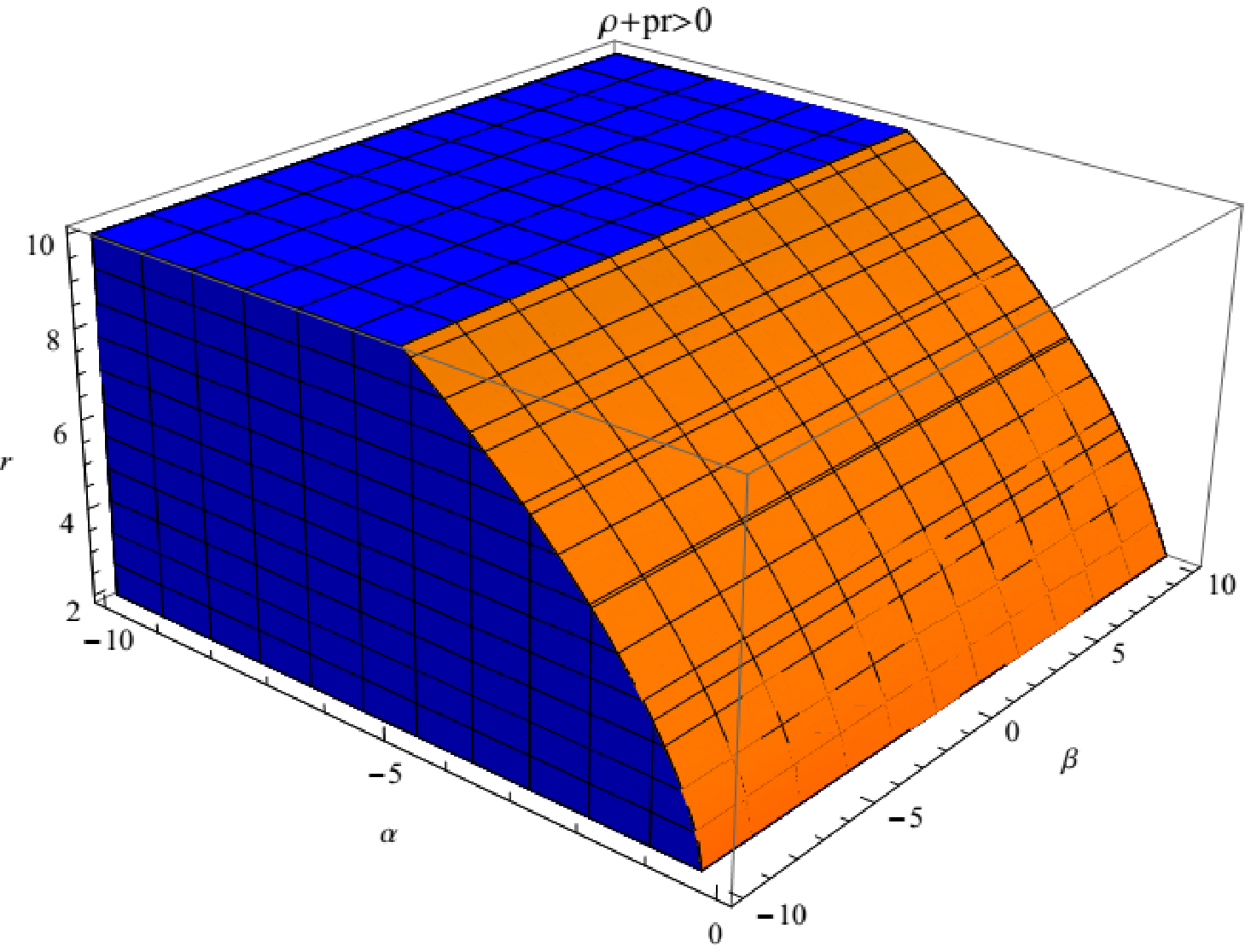,width=.48\linewidth}
\caption{The evolution of $\rho+P_r$ w.r.t. $r$, with the left side representing $\lambda>0$ and the right side representing $\lambda<0$.} \label{f2}
\end{figure}

\begin{itemize}
  \item For the validity of $\rho+P_r \geq 0$, as shown in Fig. (\ref{f2}).\\
  we see that $r\rightarrow 9$, the other parameter should be $\alpha\rightarrow -9.70$ and $\beta\rightarrow -9.50$ while for $r\rightarrow 0.1$ then $\alpha\rightarrow -5.3$ and $\beta\rightarrow 9.99$.\\

For $\alpha\rightarrow -1.89$, the other parameter should be $r\rightarrow 9$ and $\beta\rightarrow 9.99$ while for $\alpha\rightarrow -10$ then $r\rightarrow 8.87$ and $\beta\rightarrow 1.49$.\\

For $\beta\rightarrow 10$, the other parameter should be $\alpha\rightarrow -7.81$ and $r\rightarrow 1.84$ while for $\beta\rightarrow -10$ then $\alpha\rightarrow -5.62$ and $r\rightarrow 1.92$.\\
\end{itemize}

\begin{figure} \centering
\epsfig{file=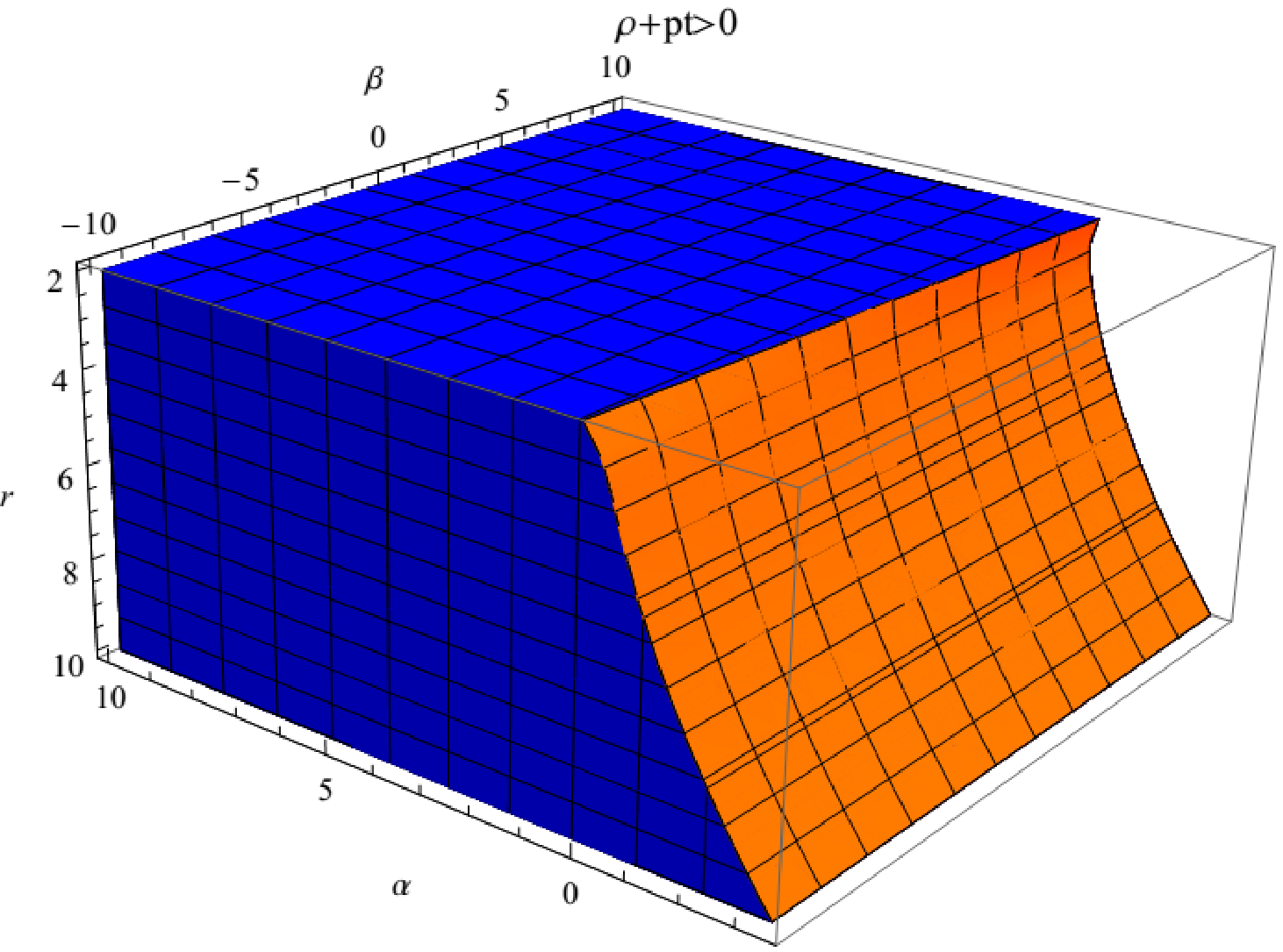,width=.48\linewidth}
\epsfig{file=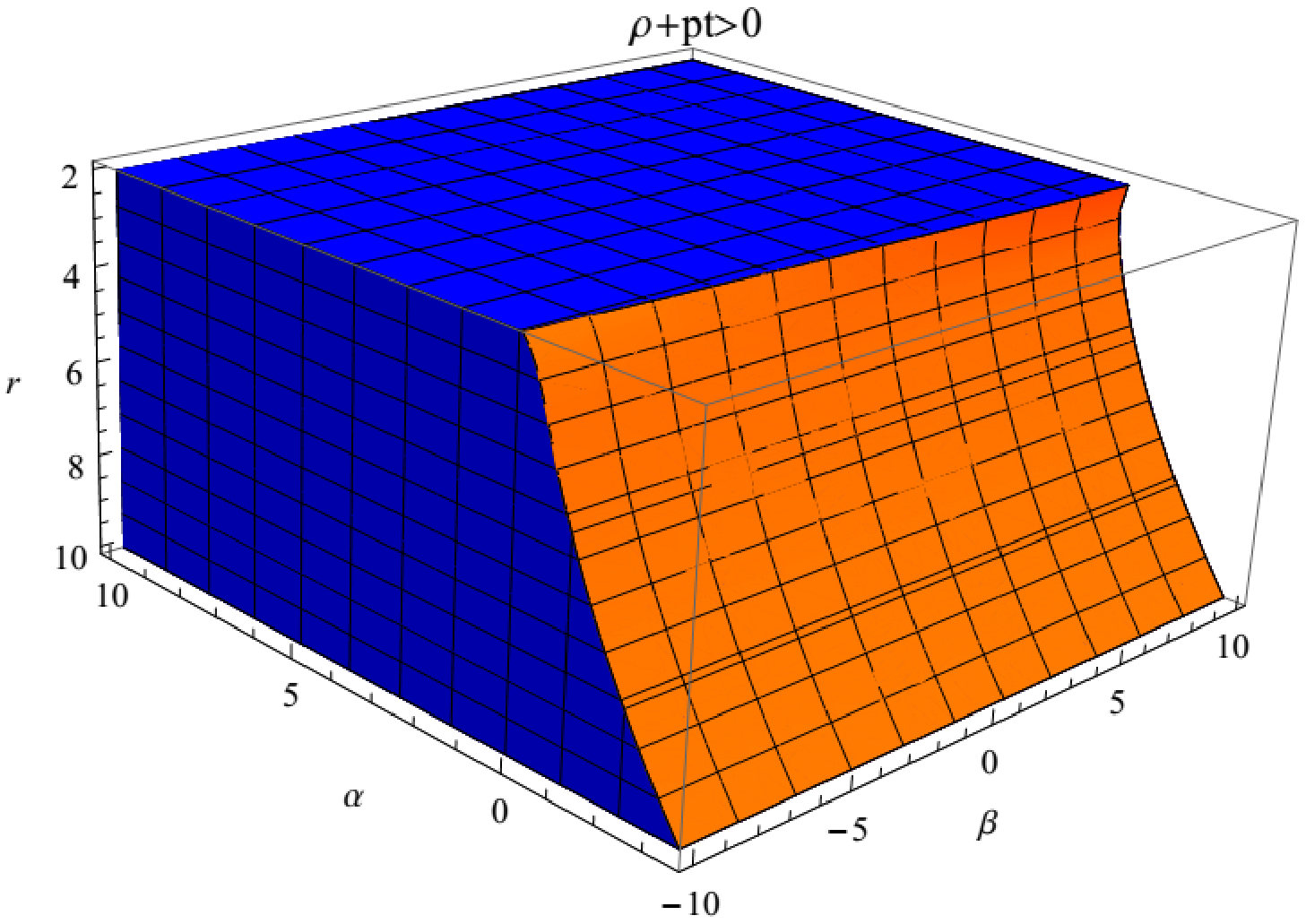,width=.48\linewidth}
\caption{The evolution of $\rho+P_t$ w.r.t. $r$, with the left side representing $\lambda>0$ and the right side representing $\lambda<0$.} \label{f3}
\end{figure}

\begin{itemize}
  \item For the validity of $\rho+P_t \geq 0$, as shown in Fig. (\ref{f3}).\\

  we see that $r\rightarrow 9$, the other parameter should be $\alpha\rightarrow 10$ and $\beta\rightarrow -10$ while for $r\rightarrow 0.1$ then $\alpha\rightarrow 9.83$ and $\beta\rightarrow 9.95$.\\

For $\alpha\rightarrow 10$, the other parameter should be $r\rightarrow 0.1$ and $\beta\rightarrow 10$ while for $\alpha\rightarrow -10$ then $r\rightarrow 9$ and $\beta\rightarrow 10$.\\

For $\beta\rightarrow 10$, the other parameter should be $\alpha\rightarrow 10$ and $r\rightarrow 8.99$ while for $\beta\rightarrow -10$ then $\alpha\rightarrow 8.75$ and $r\rightarrow 7.42$.
\end{itemize}

\begin{figure} \centering
\epsfig{file=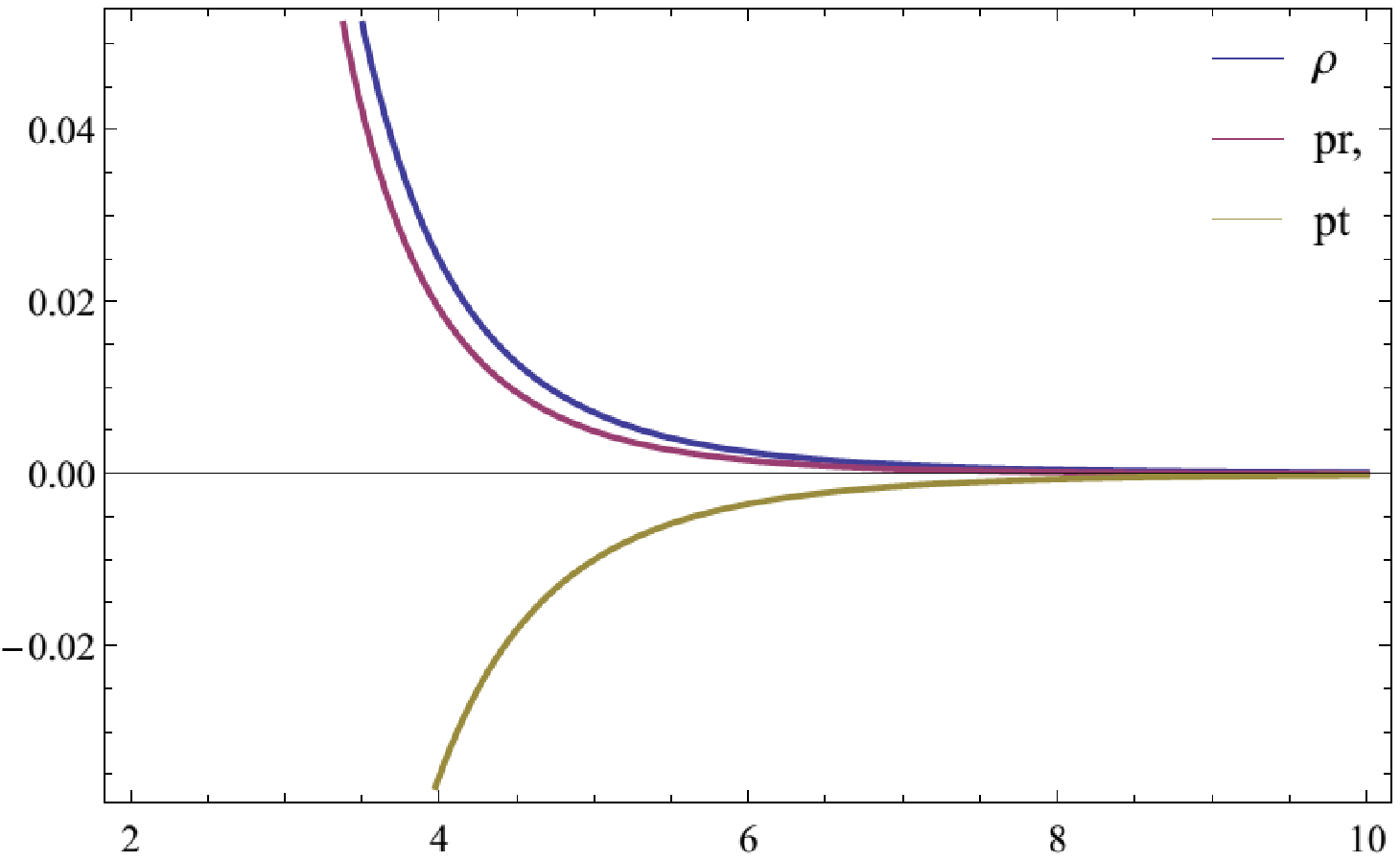,width=1\linewidth}
\caption{The behaviour of $\rho(r), P_r (r)$ and $P_t (r)$ w.r.t. $r$.} \label{f3a}
\end{figure}

While, the behavior od $\rho$, $P_r$ and $P_t$ are shown in in fig. (\ref{f3a}).

\subsubsection{A State of Equilibrium}
\begin{figure} \centering
\epsfig{file=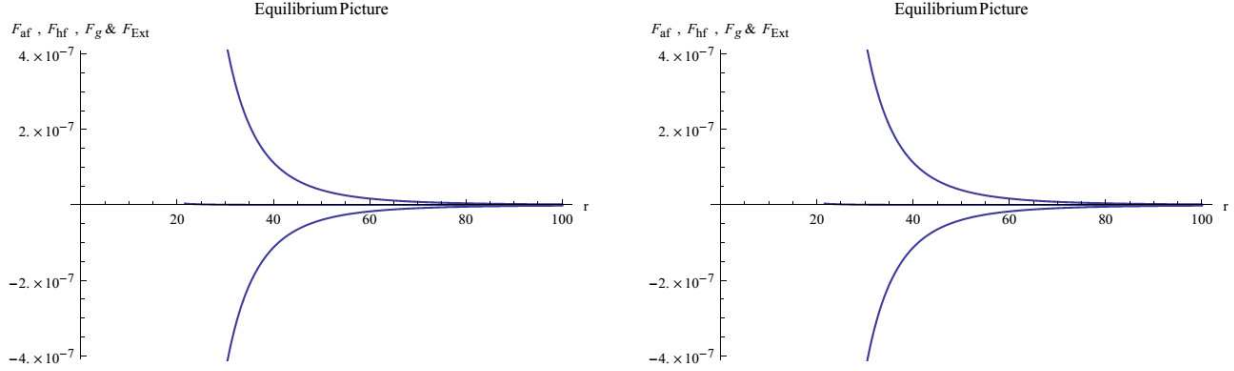,width=1\linewidth}
\caption{The behaviour of different forces, with the left side representing $\lambda>0$ and the right side representing $\lambda<0$.} \label{f4}
\end{figure}

Here, we evaluate the classification of WH-models in the equilibrium state. One can find equilibrium state for WH by solving the Tolman-Oppenheimer-Volkov (TOV) equation, given as,

\begin{equation}\label{zz1}
\frac{{d{P_r}}}{{dr}} + \frac{2}{r}\left(\rho -{P_r}\right)+a'\left(\rho  + {P_r}\right)+\frac{\lambda}{3(1+\lambda)}\frac{d}{dr}(3\rho+P_r-2P_t)= 0.
\end{equation}
The above expression (\ref{zz1}) shows that the gravitational ($F_{gf}$), the hydrostatic ($~F_{hf}$), the extra (matter-coupling) force ($F_{ext}$) and, the anisotropic ($F_{af}$) are the major forces here. The mathematical expressions of these forces are as follows:

\begin{equation}\nonumber
{F_{gf}} = - \frac{1}{2}{\sigma '[\sigma  + {P_r}]}, ~~{F_{hf}}=-\frac{{d{P_r}}}{{dr}}, ~~{F_{af}} = \frac{{2\Pi}}{r}, F_{ext}=\frac{\lambda}{3(1+\lambda)}\frac{d}{dr}(3\rho+P_r-2P_t)
\end{equation}
where $\sigma=2 a(r)$. Keeping in view the above expressions of forces, the Eq.(\ref{zz1}) may be reduced to the following,

\begin{equation}\label{zz2}
{F_{gf}}+{F_{hf}}+{F_{af}}+F_{ext}= 0.
\end{equation}
The behavior for these forces can be shown in Fig. (\ref{f4}), in which we see that the $F_g$ and $F_{ext}$ are almost negligible while the reaming two forces cancel the effects of each others.

\subsection{Isotropic Matter Distribution}
\begin{figure} \centering
\epsfig{file=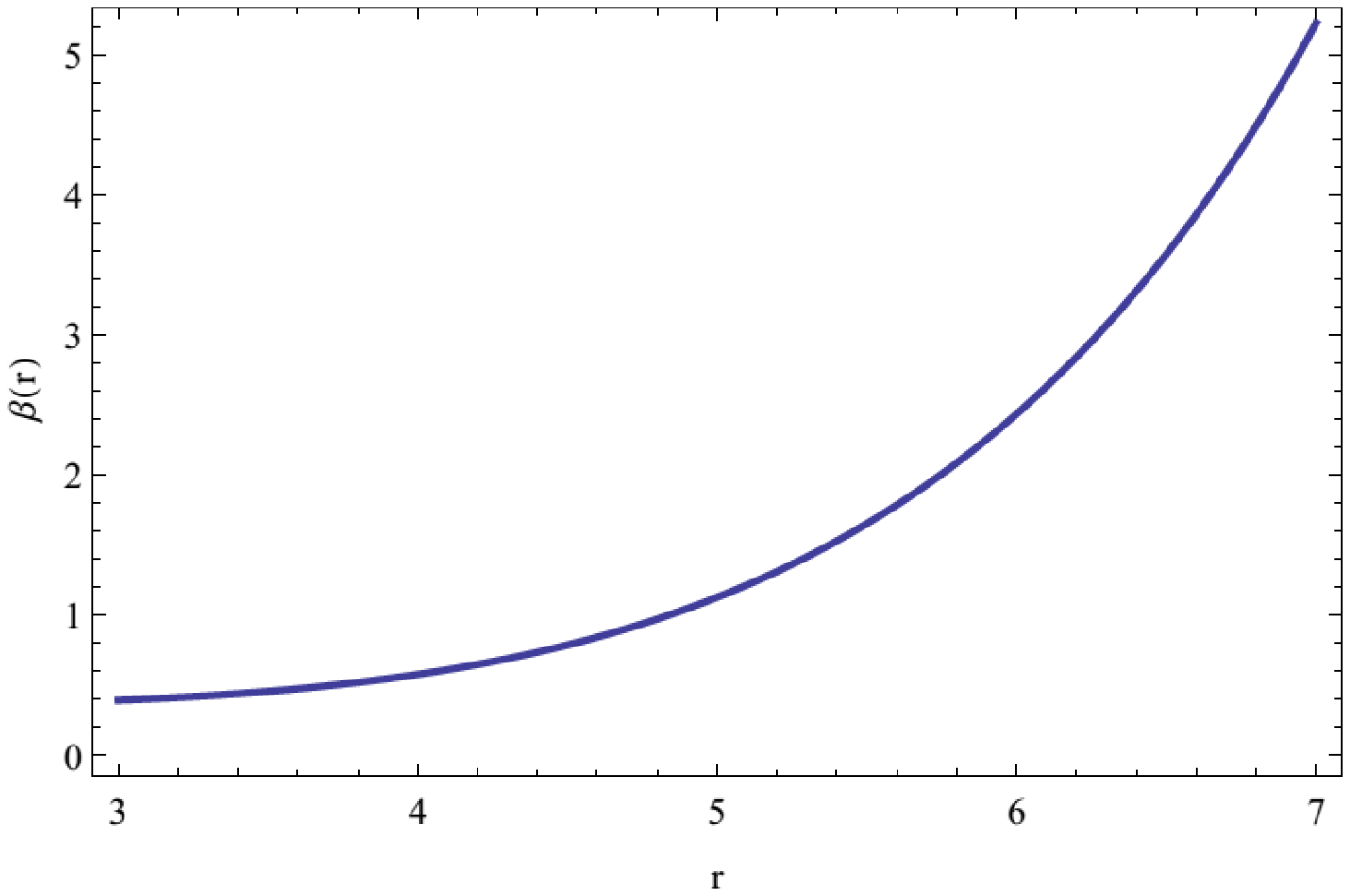,width=0.48\linewidth}
\epsfig{file=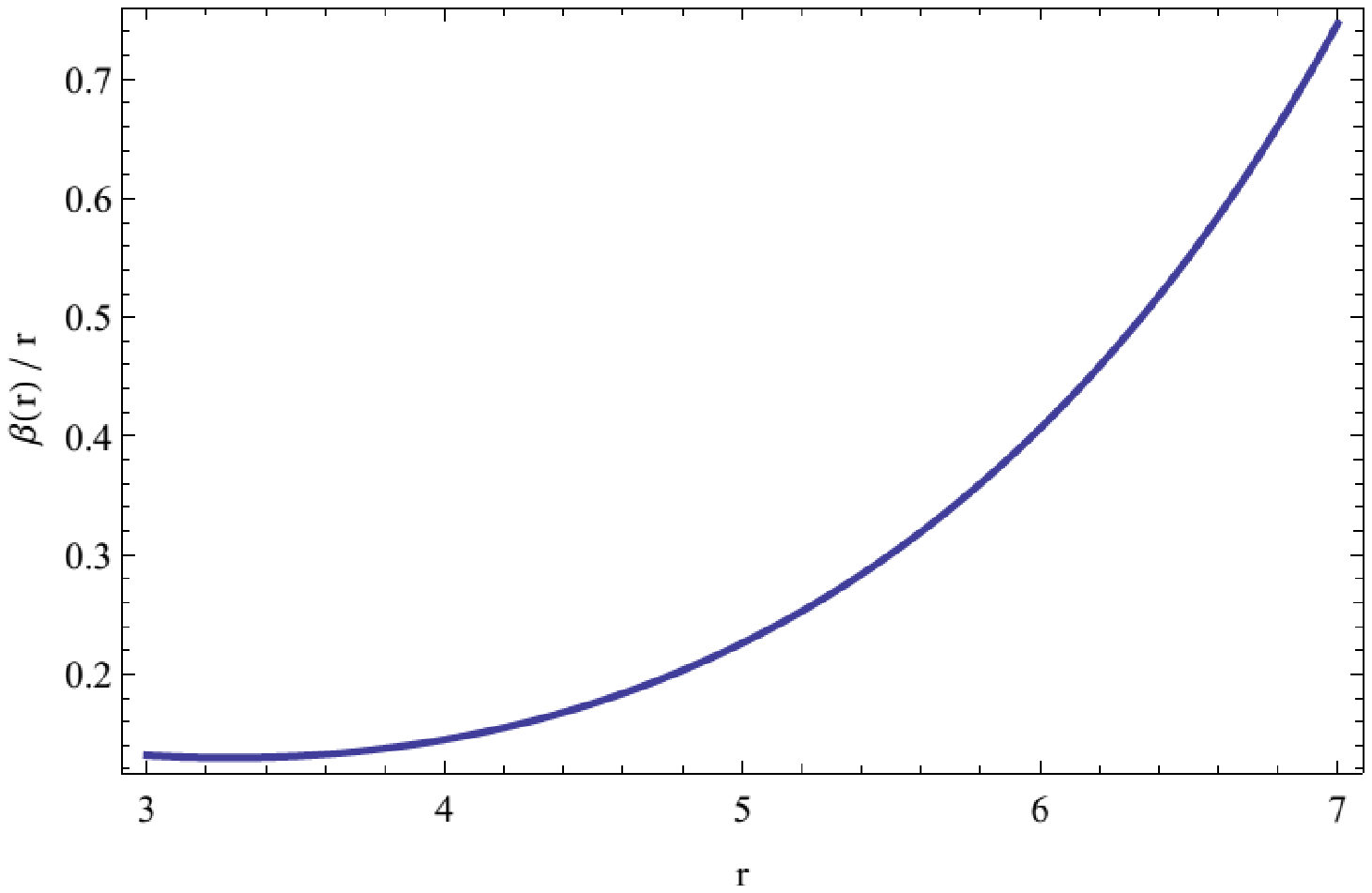,width=0.48\linewidth}
\caption{The evolution of $\beta(r)$ and $\beta(r)/r$ w.r.t. $r$ for isotropic matter distribution.} \label{f5}
\end{figure}

\begin{figure} \centering
\epsfig{file=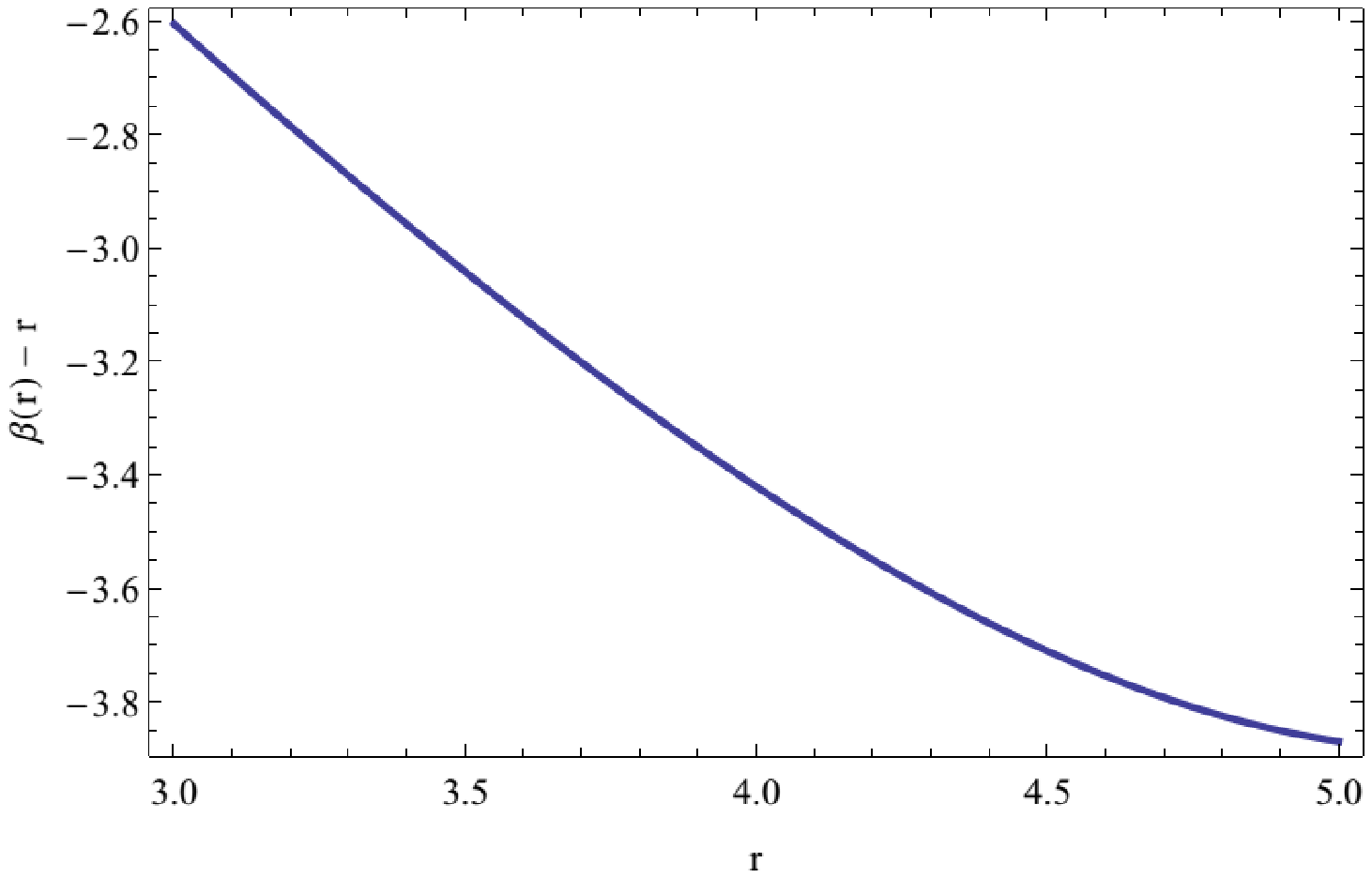,width=0.48\linewidth}
\epsfig{file=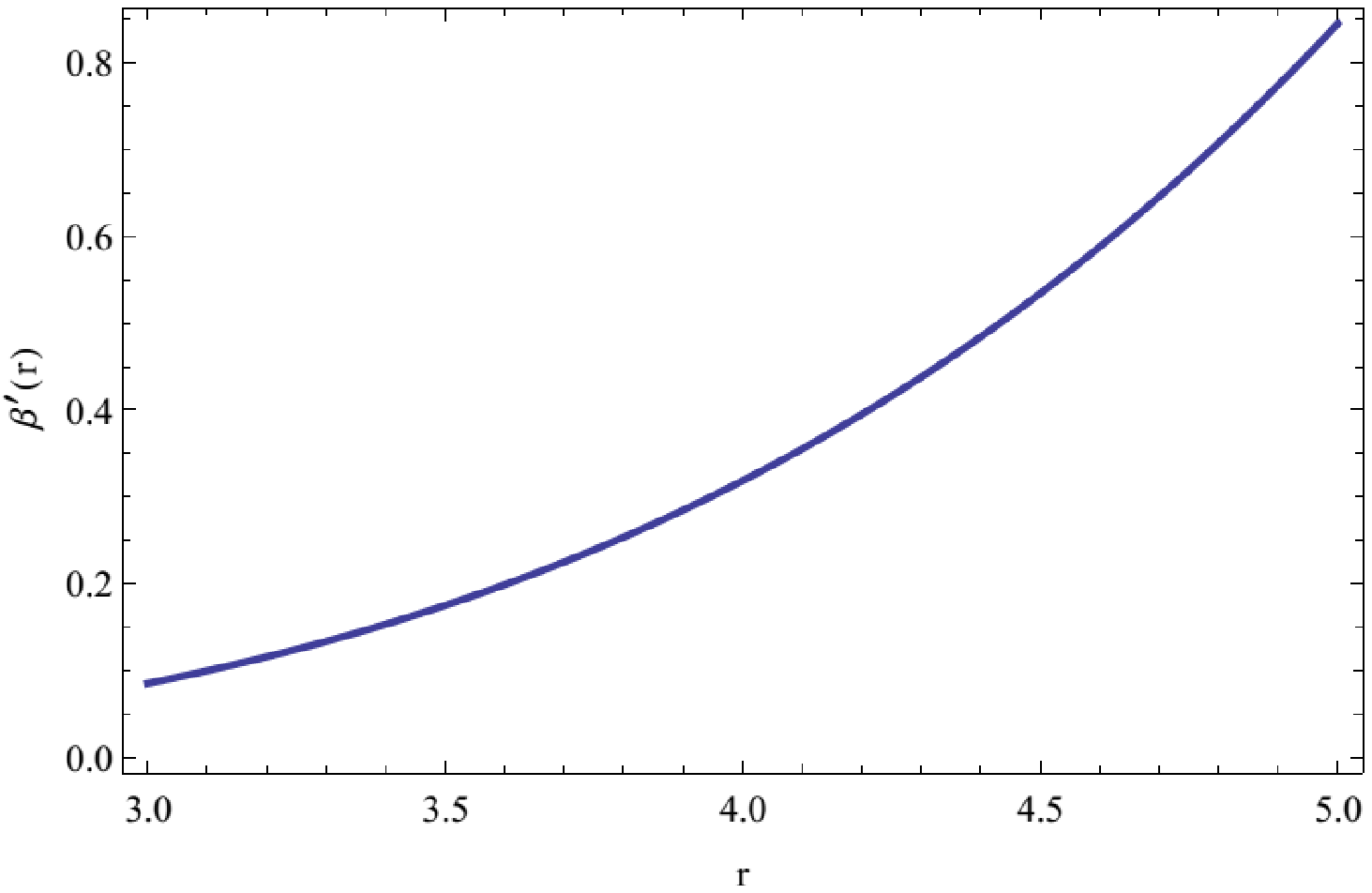,width=0.48\linewidth}
\caption{The evolution of $\beta(r)-r$ and $\beta'(r)$ w.r.t. $r$ for isotropic matter distribution.} \label{f6}
\end{figure}

\begin{figure} \centering
\epsfig{file=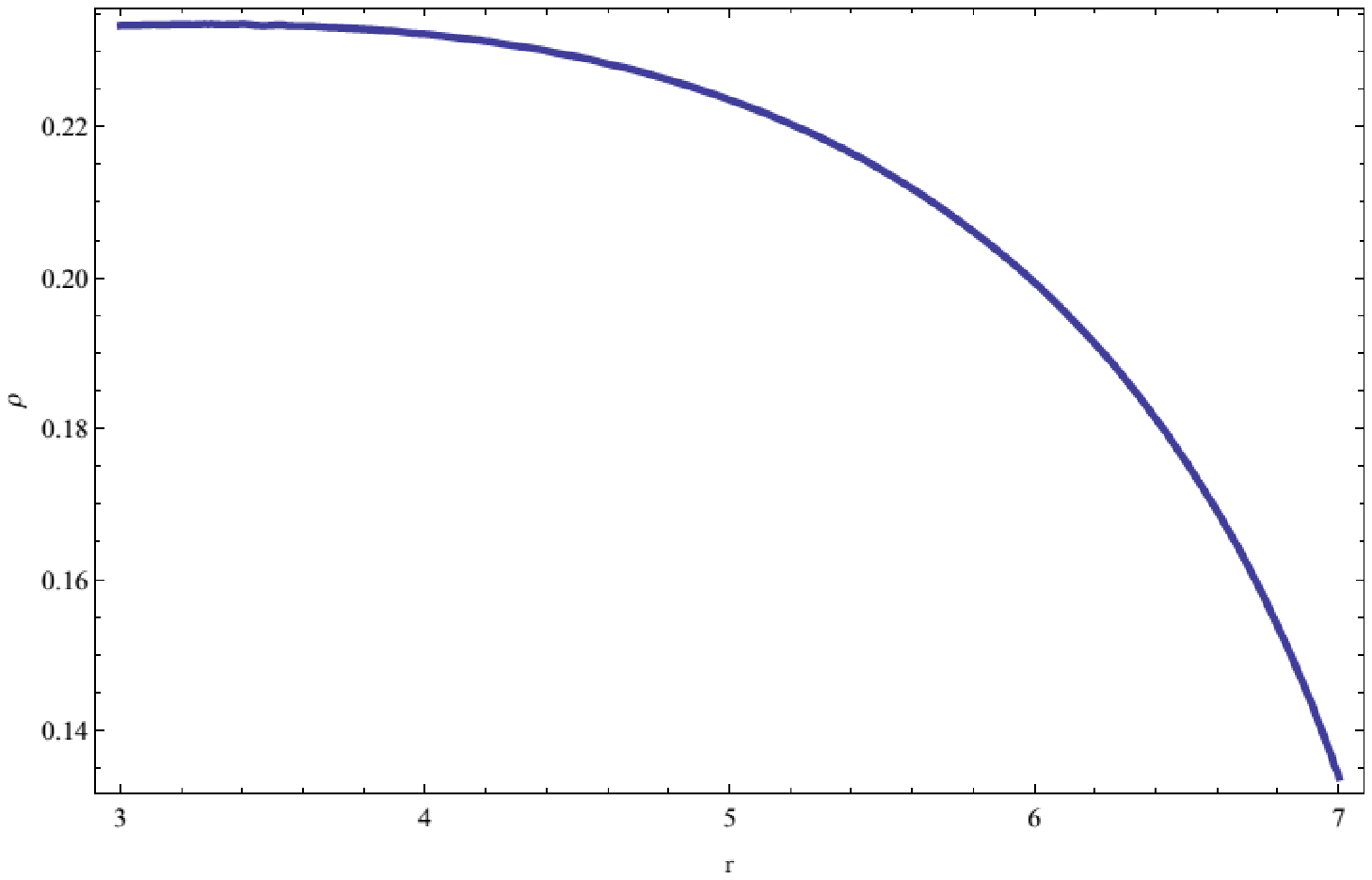,width=0.48\linewidth}
\epsfig{file=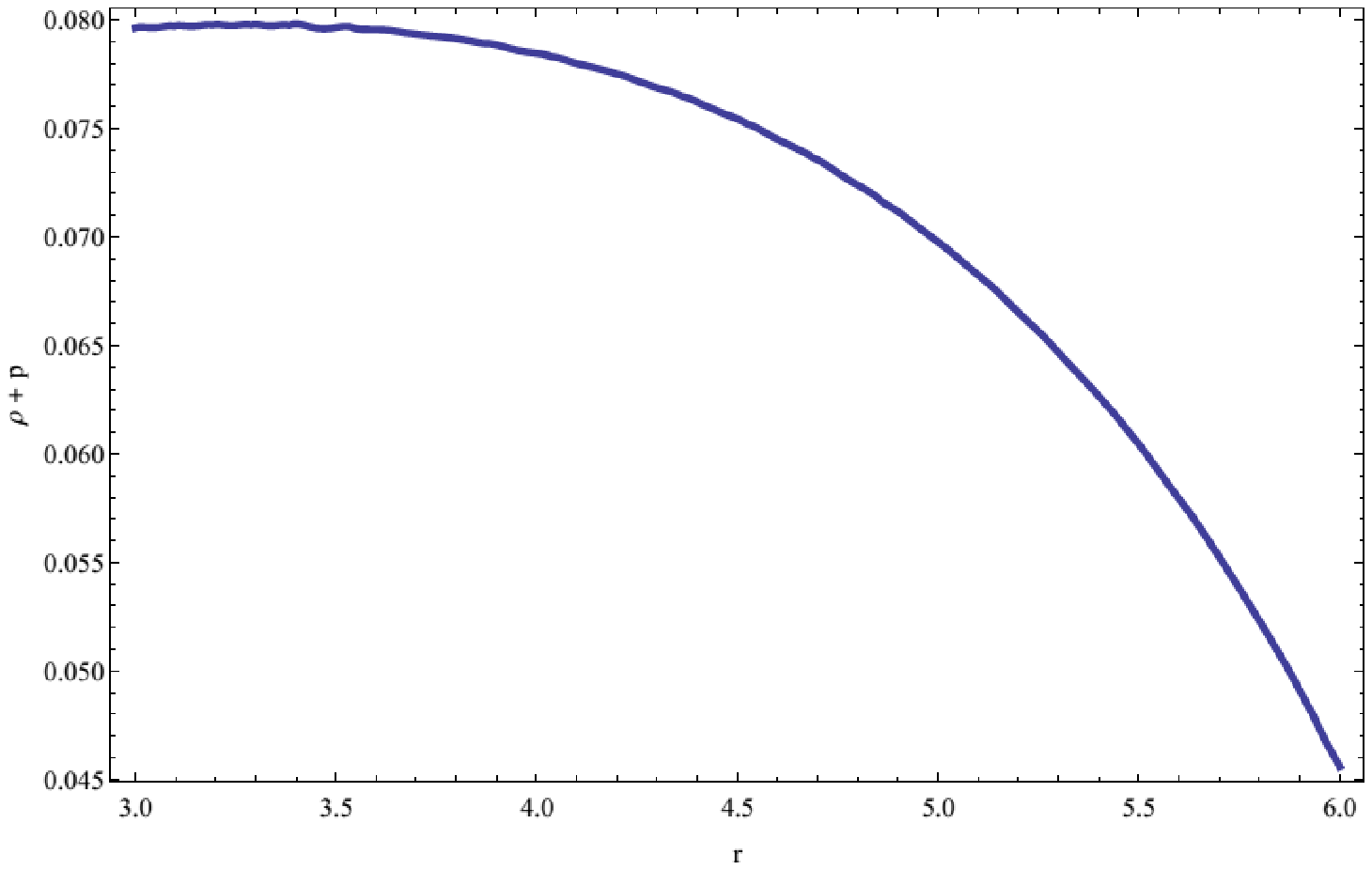,width=0.48\linewidth}
\caption{The evolution of $\rho(r)$ and $\rho+P$ w.r.t. $r$ for isotropic matter distribution.} \label{f7}
\end{figure}

In our case, we assume the coupling of a perfect matter distribution with WH geometry. Researchers in the field of relativistic astrophysics have used isotropic matter distributions to look at a number of important problems in astrophysics. Some of these problems are the gravitational collapse rate, the stability analysis of astronomical systems, the system's energy density irregularities, the universe's stable configurations maintenance and many others.
In our case, we have a system that allows for equal pressure components ($P_r = P_t =P$ or $\Pi =0$). Given the circumstances, the solutions to equations for $P_r$ and $P_t$ combine to form a non-linear third-order differential equation. Now, if we solve this equation again by incorporating $\beta(r)$ and $b(r)$, we get a third-order non-linear differential equation. We cannot use analytical methods to solve this updated third-order non-linear differential equation, so we use different numerical methods to solve it.\\
Along with the numerical techniques, we implement some initial and boundary conditions for solving this updated differential equation. The boundary conditions used are: $\beta(r_0)=03$, ${\beta^{\prime}}(r_0)=0$ and $\beta(10)=1$. Finally, we plot different solutions to this equation to check the role of NEC and WEC, as shown in Fig. (\ref{f5})-(\ref{f7}).

\subsection{Barotropic EoS}
\begin{figure} \centering
\epsfig{file=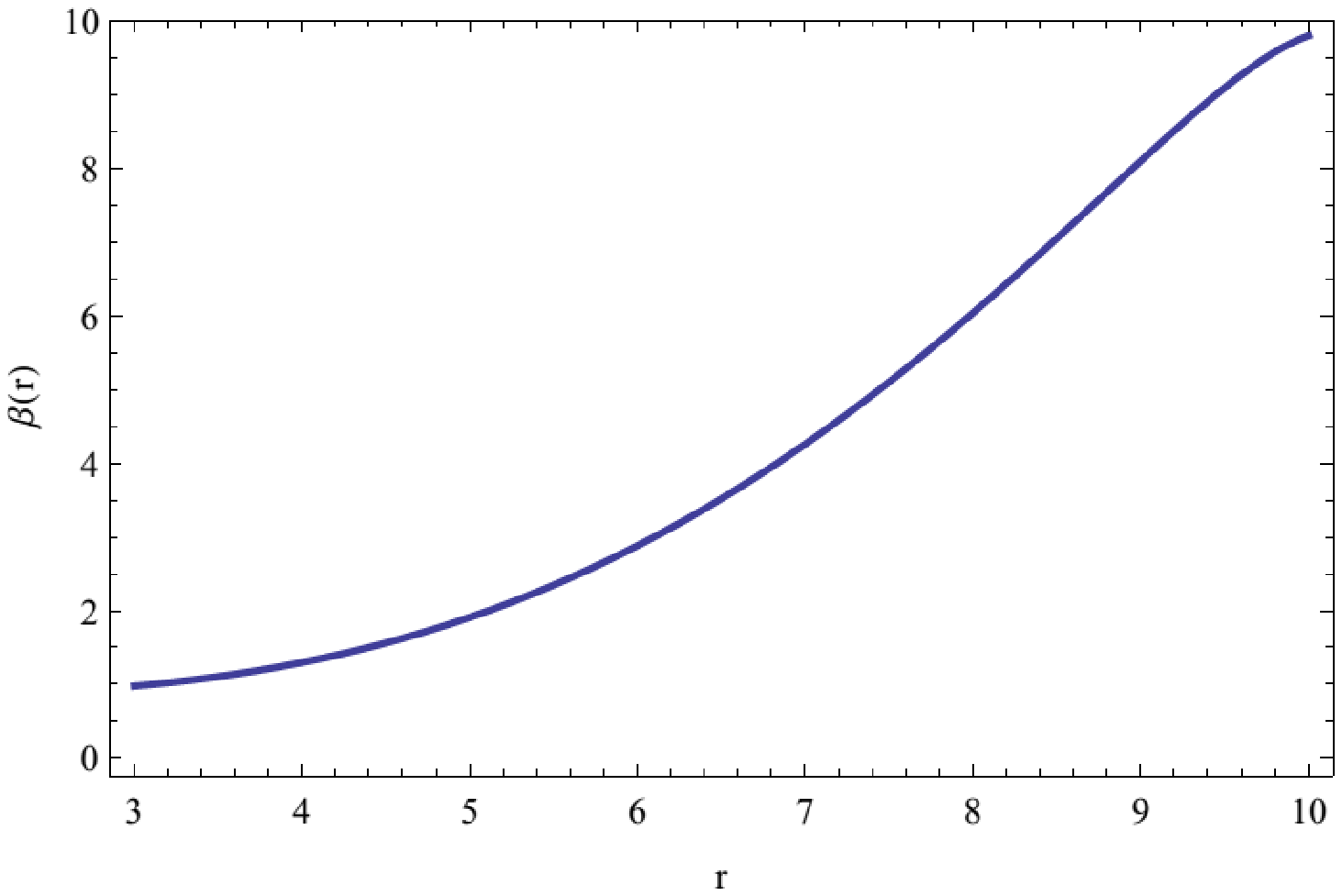,width=0.48\linewidth}
\epsfig{file=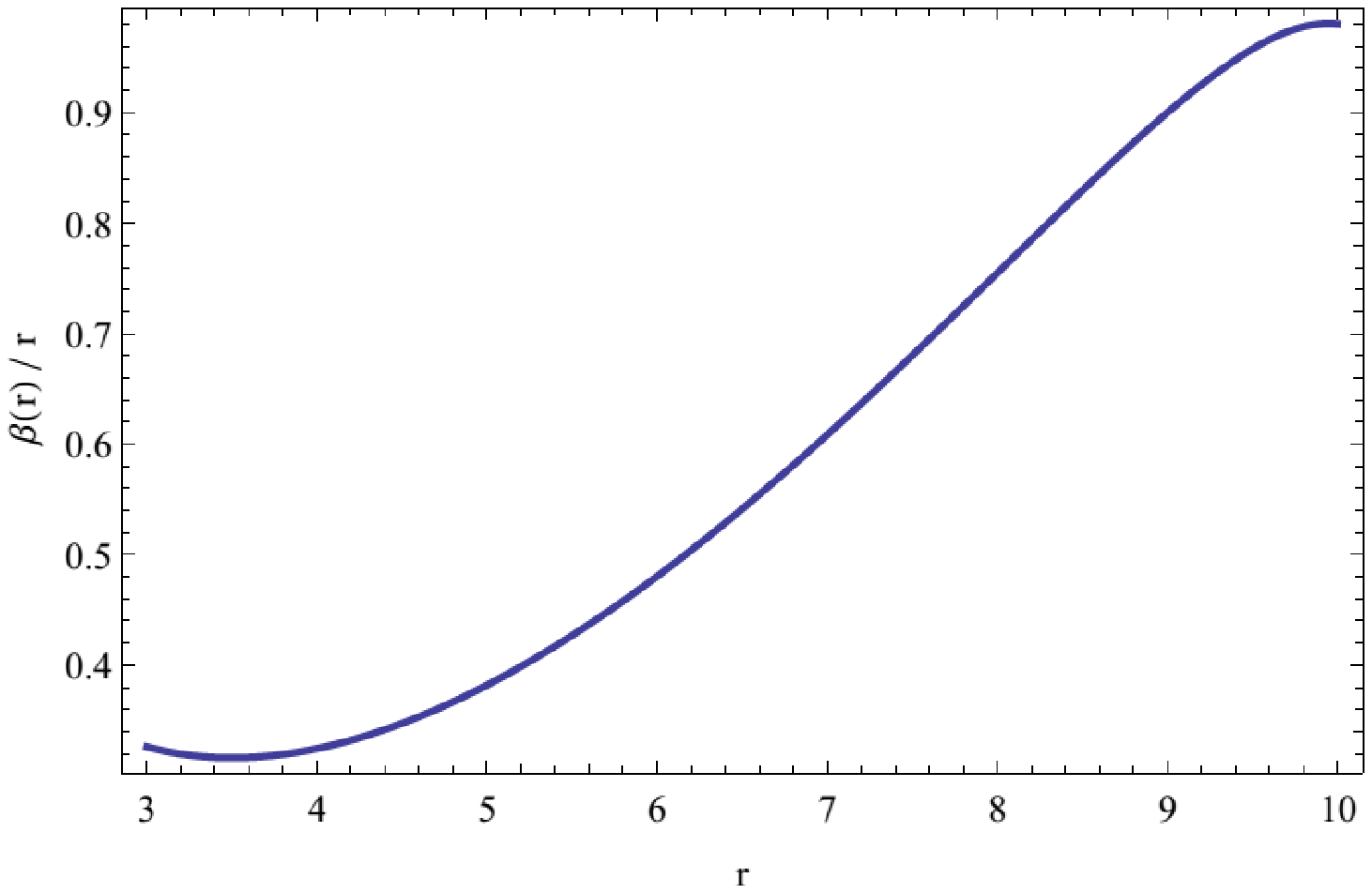,width=0.48\linewidth}
\caption{The evolution of $\beta(r)$ and $\beta(r)/r$ w.r.t. $r$ for barotropic EoS.} \label{f8}
\end{figure}

\begin{figure} \centering
\epsfig{file=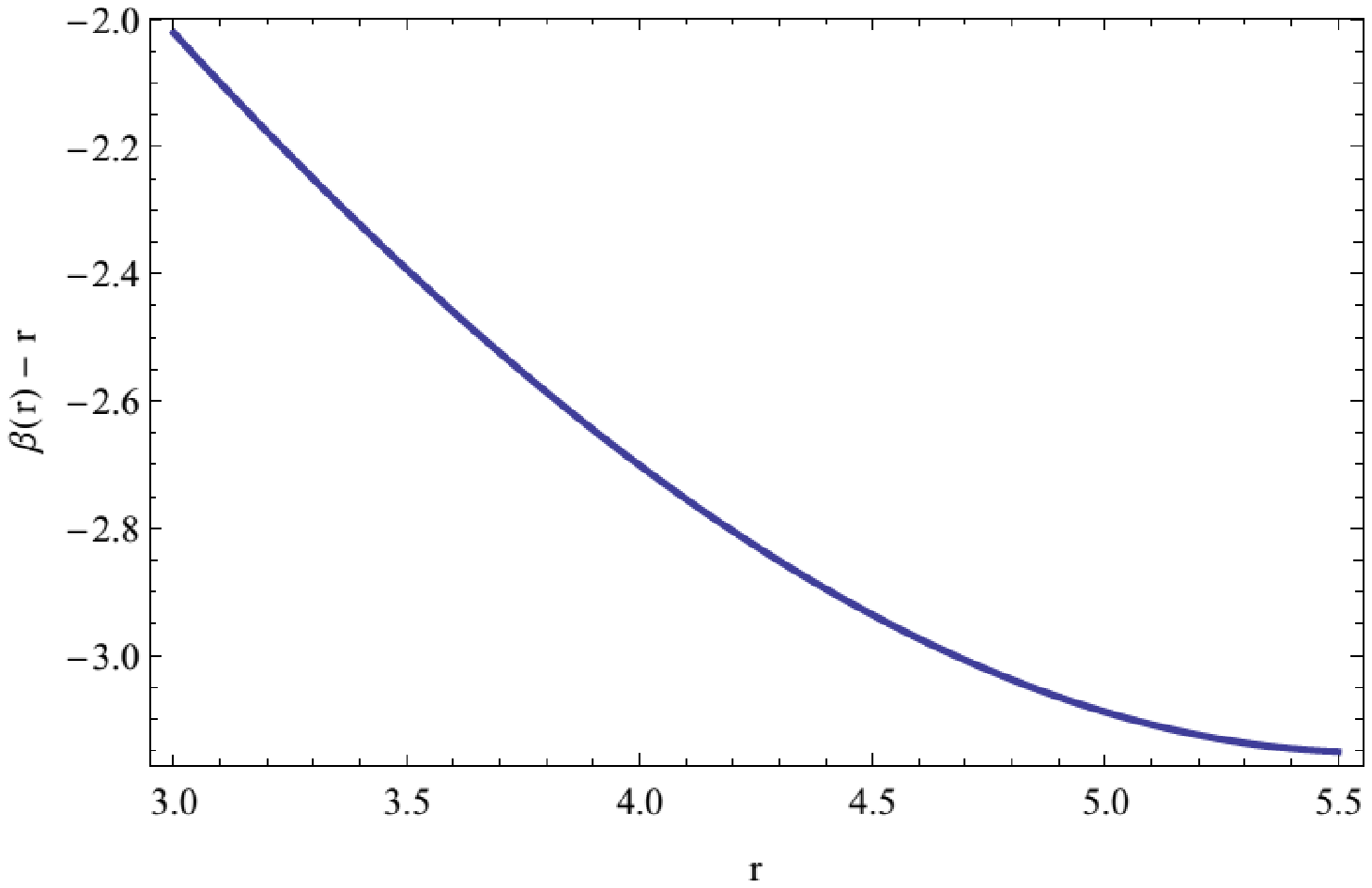,width=0.48\linewidth}
\epsfig{file=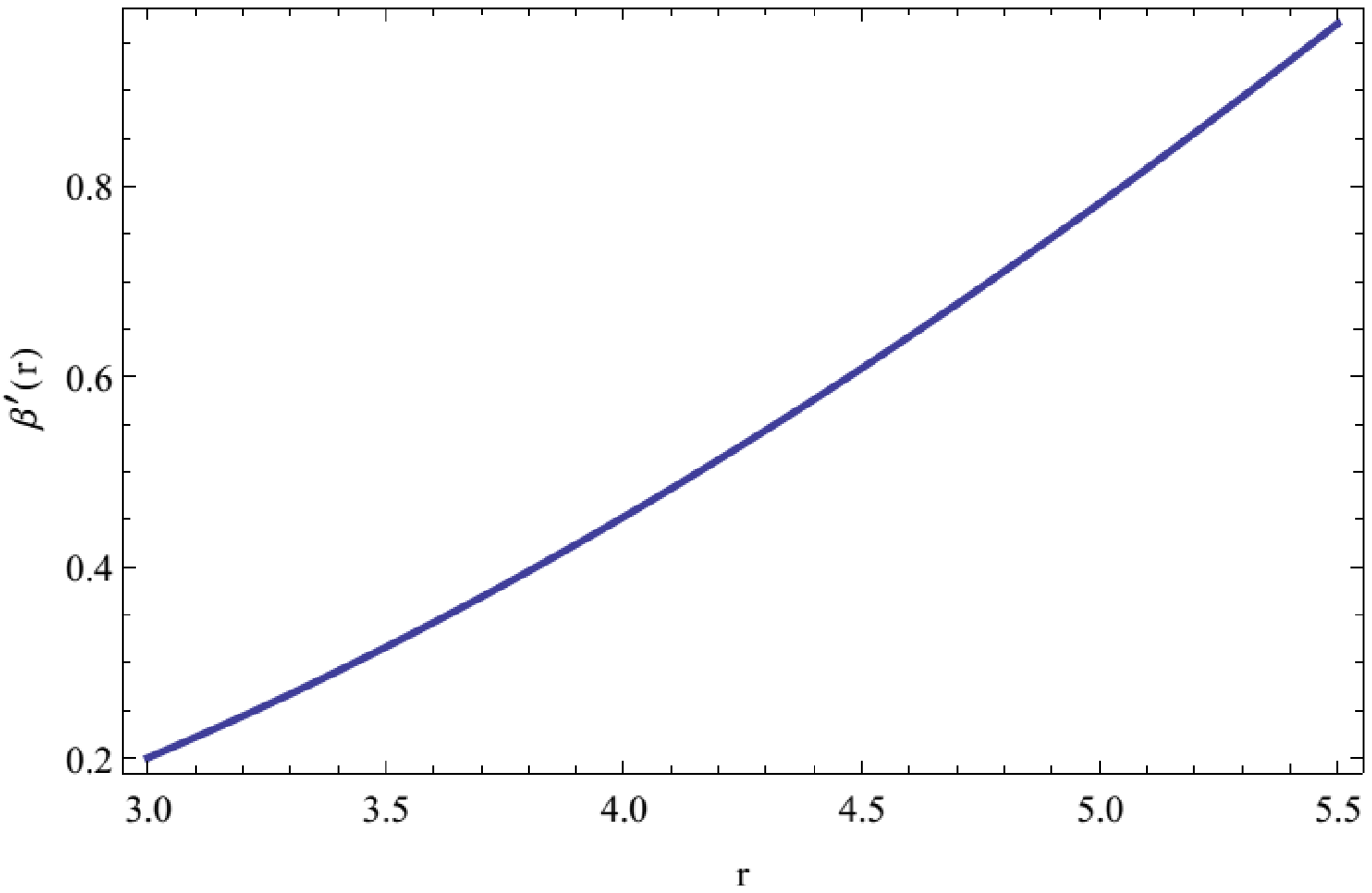,width=0.48\linewidth}
\caption{The evolution of $\beta(r)-r$ and $\beta'(r)$ w.r.t. $r$ for barotropic EoS.} \label{f9}
\end{figure}

\begin{figure} \centering
\epsfig{file=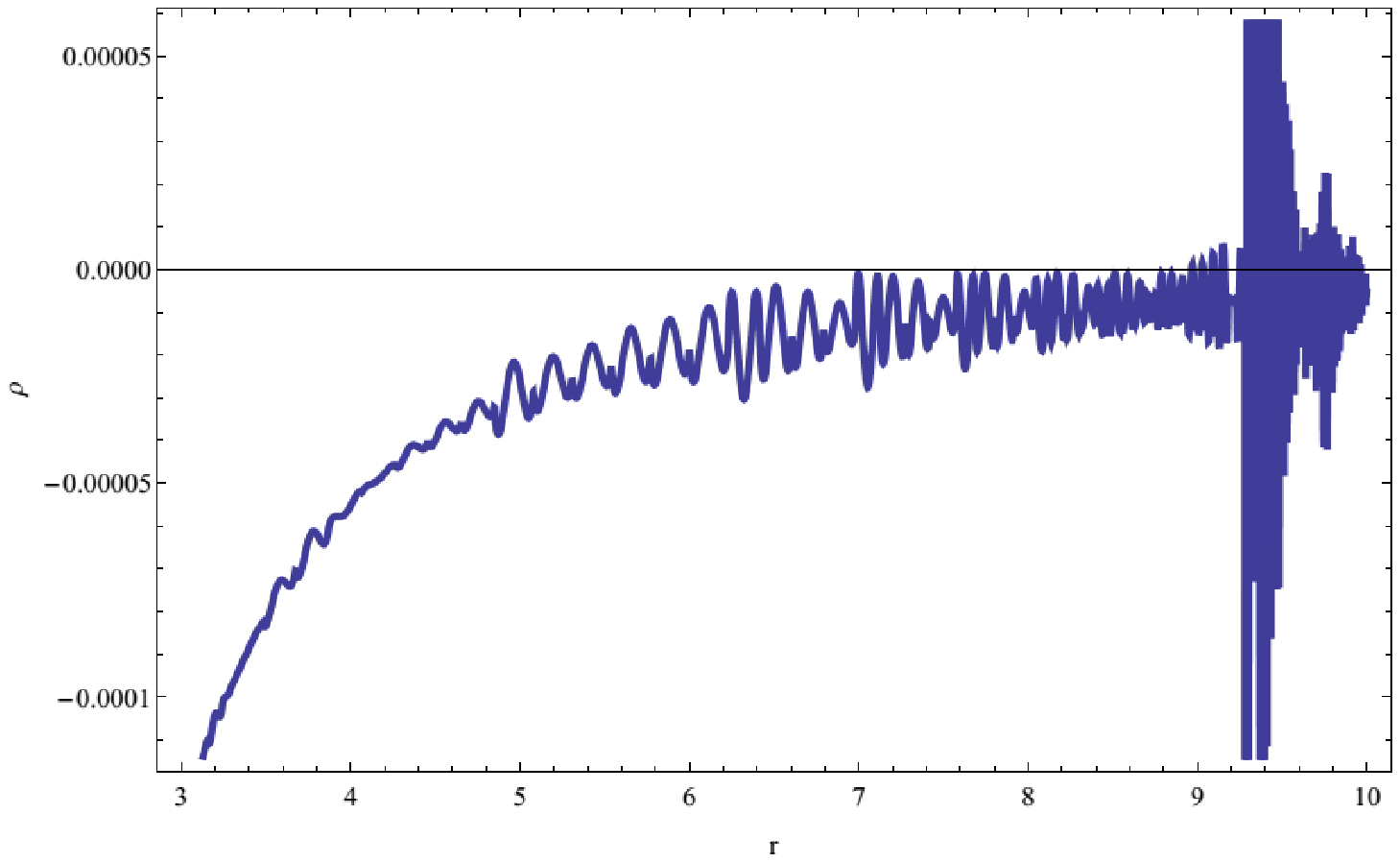,width=0.48\linewidth}
\epsfig{file=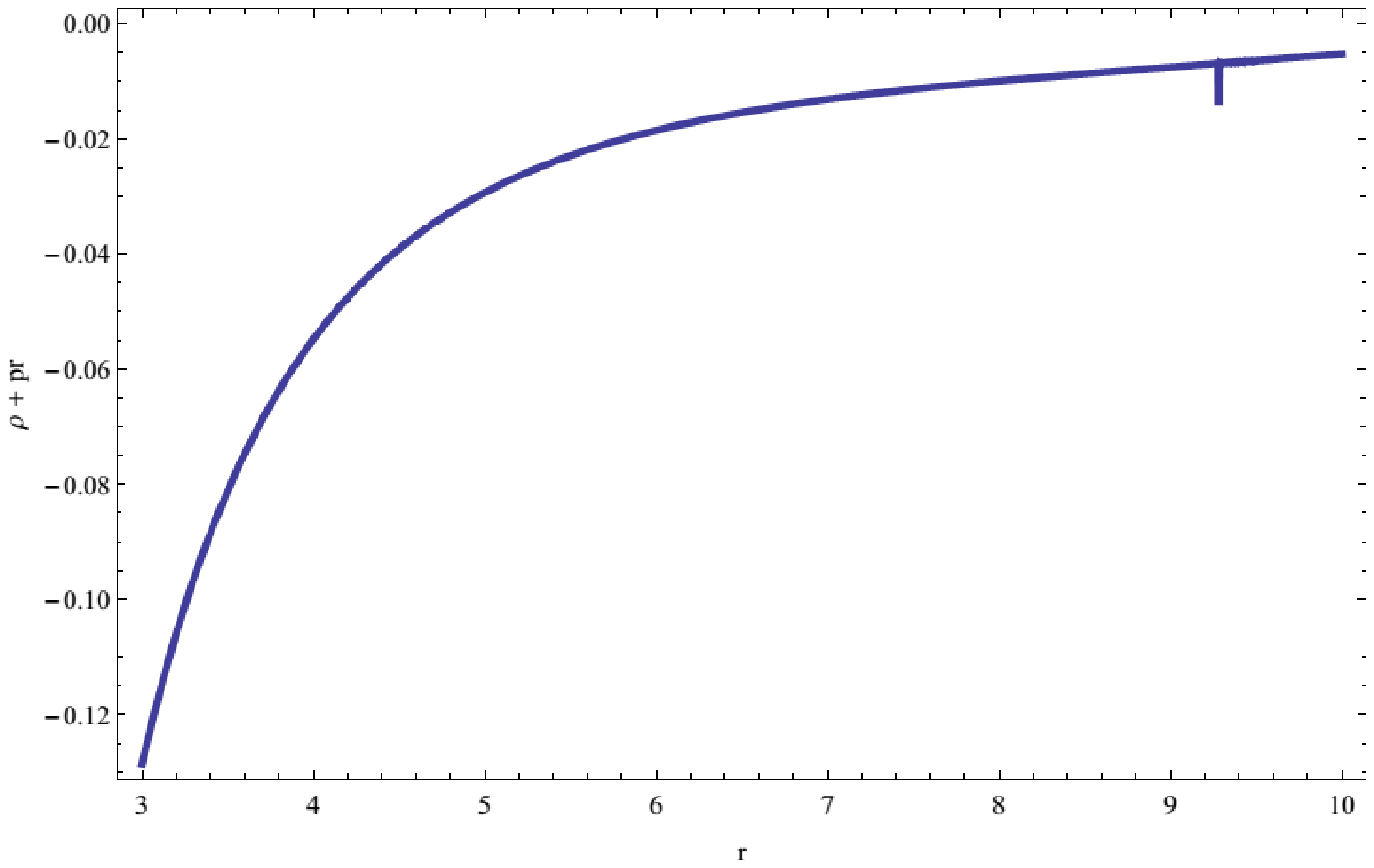,width=0.48\linewidth}
\epsfig{file=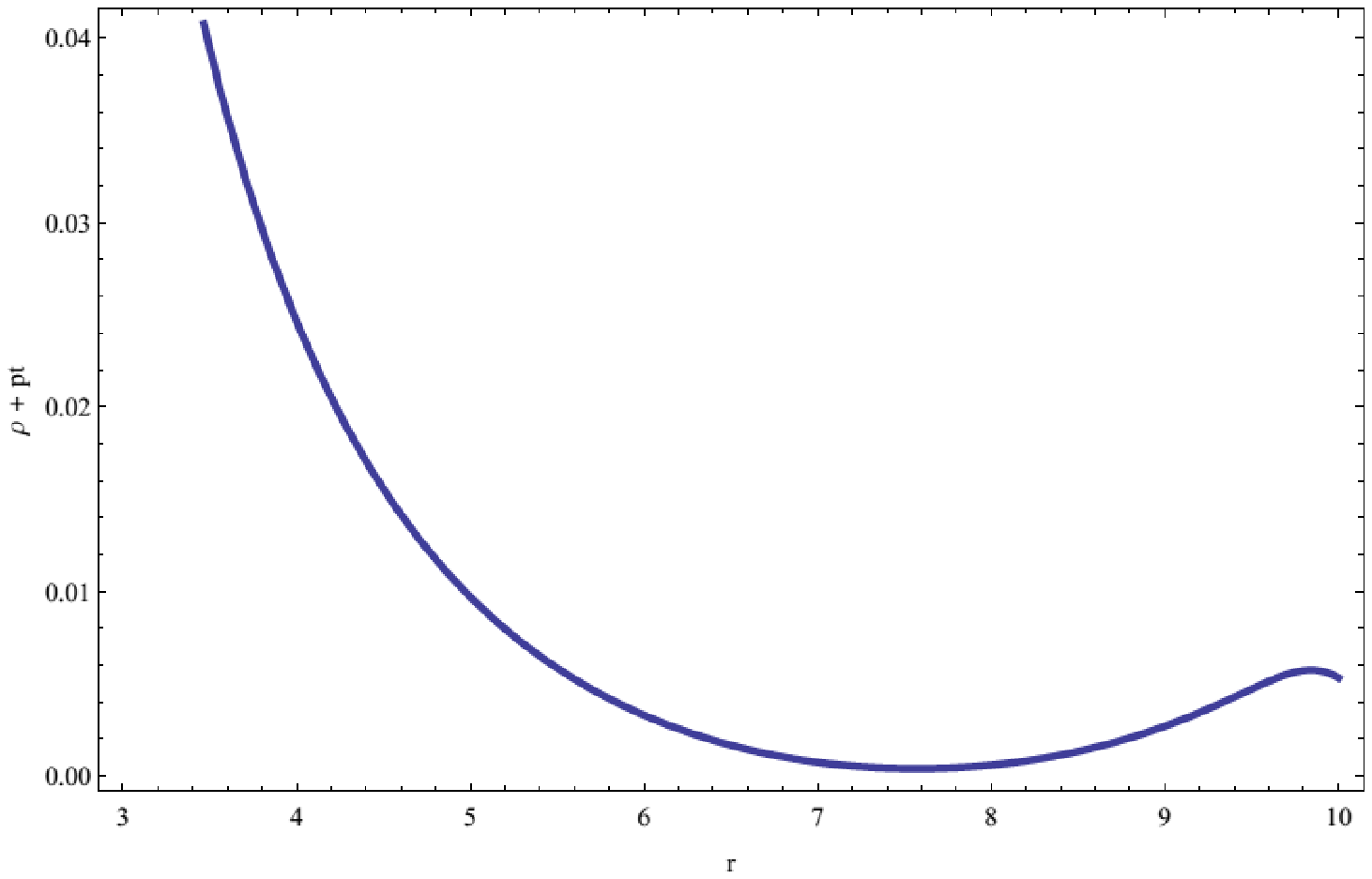,width=0.48\linewidth}
\caption{The evolution of $\rho(r)$, $\rho+P_r$ and $\rho+P_t$ w.r.t. $r$ for barotropic EoS.} \label{f10}
\end{figure}

Here, we demonstrate the relationship between $\rho $ and $P$ using a well-established cosmological-EoS and a dimensionless parameter ($k$). The realistic modeling of the barotropic-EoS results in the  representations: $P_r=k\rho$ or $P_r - k \rho = 0$. It is possible to broaden the scope of this study using equivalent $P_t$ with $\rho$ and EoS parameters (say $n$) as: $P_t=n\rho$ \cite{rref2.1}. In this case, we use the value of $P_r$ and $\rho$ along with transform $b(r)$ to $\beta(r)$ to get a third-order nonlinear differential equation. We then use numerical methods to draw different plots of $\beta(r)$ that are illustrated in Fig.\ref{f8}. In the Fig.\ref{f8}, left plot illustrates an increasing distribution of $\beta(r)$ for barotropic-EoS and that apparently WH throat lies at $r_0 =3$ where $\beta(3)=3$. The right plot of the Fig.\ref{f8} shows that $1>\beta^{\prime}(r_0)$. The flare-out and asymptotic limits can be investigated using the plots ($\frac{\beta(r)}{r}$ and $\beta(r)-r$) shown in Fig.\ref{f9}. The left plot of the Fig.\ref{f9} shows that the condition $\frac{\beta(r)}{r} \rightarrow 0$ is not valid in the range $r \rightarrow\infty$ . While the right plot of the Fig.\ref{f9} shows that the flare-out limit is applicable. The Fig.\ref{f10} shows that the WEC is not satisfied while the NEC is fulfilled. As a result, no feasible WH configurations are located in this specific region.

\section{Conclusions}
Previous research establishes that for an ideal and viable WH-model, a mandatory component is required. This component must obey a certain limit of violations of the ECs in a given feasible system. Nonetheless, considering the MGTs framework obtains the outcome of $T_{\sigma \lambda}^{\textit{effective}}$ fulfils ECs while passes through WH throat. Furthermore, the additional curvature corrections support the extraordinary
WH geometries. We used $f(R,G,T)$-MGT framework for obtaining static spherically symmetric WH solutions, that couples with isotropic, anisotropic, and barotropic EoS. We investigated the impact of $f(R,G,T)$-MGT on WH models by investigating the role of various ECs (WEC, NEC) with different matter distributions (perfect, anisotropic, and barotropic fluids). Many designated methods are proposed in the literature for describing WH models. Various schemes are used in these methods, such as considering the $\beta(r)$ for WH and then investigating ECs, selecting appropriate fluid mechanisms and then computing $\beta(r)$. Our investigations in this paper are limited to the framework of $f(R,G,T)$-MGT, which includes matter couplings with Ricci variables. We used the following model to examine the viability of WH solutions: $f(R,G,T) = R +\alpha R^2+\beta G^n+\gamma G\ln(G)+\lambda T$. Our calculations are very tedious and involve comparatively complicated nonlinear differential equations with unknown functions ($\rho, P_r, P_t, a, b, f(R,G,T)$), so some unique factors play a crucial role in these calculations.\\
Considering our findings, we report that the violation of NEC by matter passing through the WH throat can be validated using the stress energy tensor. Next to this, we briefly studied solutions of the gravitational FEqs using different matter distributions (isotropic, anisotropic, and barotropic fluids). We studied these equations using $\beta(r)$ with quadratic Ricci scalar function and assuming $f(T)=\lambda T$. In the case of isotropic and barotropic matter distributions, the nonlinear differential equations are computed using numerical methods and the corresponding results are presented by plotting $\beta(r)$. Our plots locate some regions in which WH fulfills the ECs. Moreover, we analyzed the significance of the equilibrium state and obtained that $F_{hf}$ and $F_{gf}$ balance each other.\\
To obtain viable WH solutions in $f(R,G,T)$-MGT, a specific model is considered, and then FEqs are determined using a set of $\beta(r)$. The behaviour of ECs (NEC, WEC) is examined by plotting different regions corresponding to a set of values of $r, \alpha$ and $\beta$. Resultantly, we obtained some regions for viable WHs in the absence of exotic matter. We also examined the equilibrium state for anisotropic matter distribution using these plots and found that $F_{hf}$ and $F_{af}$ cancel the effect of each other while the $F_g$ and $F_{ext}$ are almost negligible. These plots shows the numerical results for $\beta(r)$ in the background of barotropic EoS.

\section*{Author Contributions:} M.I, A.R.A., F.K., N.G.,  H.I.A., K.S.N., A.-H.A.-A. have contributed equally. All
authors have read and agreed to the published version of the manuscript.
\section*{Funding:} This work was supported by Princess Nourah bint Abdulrahman University Researchers
Supporting Project number (PNURSP2023R106), Princess Nourah bint Abdulrahman University,
Riyadh, Saudi Arabia.

\section*{Data availability statement}

All data generated or analyzed during this study are included in this published article.

\section*{Acknowledgments:} This work was supported by Princess Nourah bint Abdulrahman University
Researchers Supporting Project number (PNURSP2023R106), Princess Nourah bint Abdulrahman
University, Riyadh, Saudi Arabia. We would like to thank the reviewers for their thoughtful comments
and efforts towards improving our manuscript.
\section*{Conflicts of Interest:} The authors declare no conflict of interest.

\vspace{0.5cm}

\bibliographystyle{unsrt}
\bibliography{mybib}
\end{document}